\documentclass[manyauthors,nocleardouble,COMPASS]{cernphprep}

\usepackage{bm}
\usepackage{color}

\usepackage{amssymb}
\usepackage{amsmath}
\usepackage{amsbsy}
\usepackage{times}
\usepackage{epsfig}
\usepackage{colordvi}
\usepackage{graphicx}
\usepackage{wrapfig,rotating}
\usepackage[numbers, square, comma, sort&compress]{natbib}
\usepackage{hyperref} 
\usepackage{multicol}
\usepackage{booktabs}
\usepackage{multirow}
\usepackage{lineno}
\RequirePackage[T1]{fontenc}
 
\makeatletter

\DeclareSymbolFont{letters}     {OML}{cmm}{m}{it}
\DeclareSymbolFont{symbols}     {OMS}{cmsy}{m}{n}
\DeclareSymbolFont{largesymbols}{OMX}{cmex}{m}{n}

\newcommand{\eea}{\end{eqnarray}}
\newcommand{\bea}{\begin{eqnarray}}
\newcommand{\beq}{\begin{equation}}
\newcommand{\eeq}{\end{equation}}
\newcommand{\bit}{\begin{itemize}}
\newcommand{\eit}{\end{itemize}}
\newcommand{\bef}{\begin{figure}}
\newcommand{\eef}{\end{figure}}
\newcommand{\bec}{\begin{center}}
\newcommand{\eec}{\end{center}}
\newcommand{\mrf}[1]{\mbox{$\mathrm{#1}$}} 

\begin{document}

\begin{titlepage}

\PHnumber{2014--213}
\PHdate{19 August 2014}

\title{Collins and Sivers asymmetries in muonproduction of pions and kaons off
transversely polarised protons}

\Collaboration{The COMPASS Collaboration}
\ShortAuthor{The COMPASS Collaboration}

\begin{abstract}
Measurements of the Collins and Sivers asymmetries for charged pions and charged
and neutral kaons produced in semi-inclusive deep-inelastic scattering of high
energy muons off transversely polarised protons are presented. The results were
obtained using all the available COMPASS proton data, which were taken in the years
2007 and 2010. The Collins asymmetries exhibit in the valence region a
non-zero signal for pions and there are hints of non-zero
signal also for kaons. The Sivers asymmetries are found to be
positive for positive pions and kaons and compatible with zero otherwise.
\end{abstract}
\vfill
\Submitted{(to be submitted to Phys. Lett. B)}

\end{titlepage}

{\pagestyle{empty}
%
%

\section*{The COMPASS Collaboration}
\label{app:collab}
\renewcommand\labelenumi{\textsuperscript{\theenumi}~}
\renewcommand\theenumi{\arabic{enumi}}
\begin{flushleft}
C.~Adolph\Irefn{erlangen},
R.~Akhunzyanov\Irefn{dubna}, 
M.G.~Alexeev\Irefn{turin_u},
G.D.~Alexeev\Irefn{dubna}, 
A.~Amoroso\Irefnn{turin_u}{turin_i},
V.~Andrieux\Irefn{saclay},
V.~Anosov\Irefn{dubna}, 
A.~Austregesilo\Irefnn{cern}{munichtu},
B.~Bade{\l}ek\Irefn{warsawu},
F.~Balestra\Irefnn{turin_u}{turin_i}
% ...
\end{flushleft}

%
%

\begin{Authlist}
\item \Idef{bielefeld}{Universit\"at Bielefeld, Fakult\"at f\"ur Physik, 33501 Bielefeld, Germany\Arefs{f}}
\item \Idef{bochum}{Universit\"at Bochum, Institut f\"ur Experimentalphysik, 44780 Bochum, Germany\Arefs{f}\Arefs{ll}}
\item \Idef{bonniskp}{Universit\"at Bonn, Helmholtz-Institut f\"ur  Strahlen- und Kernphysik, 53115 Bonn, Germany\Arefs{f}}
\item \Idef{bonnpi}{Universit\"at Bonn, Physikalisches Institut, 53115 Bonn, Germany\Arefs{f}}
\item \Idef{brno}{Institute of Scientific Instruments, AS CR, 61264 Brno, Czech Republic\Arefs{g}}
\item \Idef{calcutta}{Matrivani Institute of Experimental Research \& Education, Calcutta-700 030, India\Arefs{h}}
\item \Idef{dubna}{Joint Institute for Nuclear Research, 141980 Dubna, Moscow region, Russia\Arefs{i}}
\item \Idef{erlangen}{Universit\"at Erlangen--N\"urnberg, Physikalisches Institut, 91054 Erlangen, Germany\Arefs{f}}
\item \Idef{freiburg}{Universit\"at Freiburg, Physikalisches Institut, 79104 Freiburg, Germany\Arefs{f}\Arefs{ll}}
\item \Idef{cern}{CERN, 1211 Geneva 23, Switzerland}
\item \Idef{liberec}{Technical University in Liberec, 46117 Liberec, Czech Republic\Arefs{g}}
\item \Idef{lisbon}{LIP, 1000-149 Lisbon, Portugal\Arefs{j}}
\item \Idef{mainz}{Universit\"at Mainz, Institut f\"ur Kernphysik, 55099 Mainz, Germany\Arefs{f}}
\item \Idef{miyazaki}{University of Miyazaki, Miyazaki 889-2192, Japan\Arefs{k}}
\item \Idef{moscowlpi}{Lebedev Physical Institute, 119991 Moscow, Russia}
\item \Idef{munichlmu}{Ludwig-Maximilians-Universit\"at M\"unchen, Department f\"ur Physik, 80799 Munich, Germany\Arefs{f}\Arefs{l}}
\item \Idef{munichtu}{Technische Universit\"at M\"unchen, Physik Department, 85748 Garching, Germany\Arefs{f}\Arefs{l}}
\item \Idef{nagoya}{Nagoya University, 464 Nagoya, Japan\Arefs{k}}
\item \Idef{praguecu}{Charles University in Prague, Faculty of Mathematics and Physics, 18000 Prague, Czech Republic\Arefs{g}}
\item \Idef{praguectu}{Czech Technical University in Prague, 16636 Prague, Czech Republic\Arefs{g}}
\item \Idef{protvino}{State Scientific Center Institute for High Energy Physics of National Research Center `Kurchatov Institute', 142281 Protvino, Russia}
\item \Idef{saclay}{CEA IRFU/SPhN Saclay, 91191 Gif-sur-Yvette, France\Arefs{ll}}
\item \Idef{telaviv}{Tel Aviv University, School of Physics and Astronomy, 69978 Tel Aviv, Israel\Arefs{m}}
\item \Idef{triest_u}{University of Trieste, Department of Physics, 34127 Trieste, Italy}
\item \Idef{triest_i}{Trieste Section of INFN, 34127 Trieste, Italy}
\item \Idef{triest_ictp}{Abdus Salam ICTP, 34151 Trieste, Italy}
\item \Idef{turin_u}{University of Turin, Department of Physics, 10125 Turin, Italy}
\item \Idef{turin_p}{University of Eastern Piedmont, 15100 Alessandria, Italy}
\item \Idef{turin_i}{Torino Section of INFN, 10125 Turin, Italy}
\item \Idef{warsaw}{National Centre for Nuclear Research, 00-681 Warsaw, Poland\Arefs{n} }
\item \Idef{warsawu}{University of Warsaw, Faculty of Physics, 00-681 Warsaw, Poland\Arefs{n} }
\item \Idef{warsawtu}{Warsaw University of Technology, Institute of Radioelectronics, 00-665 Warsaw, Poland\Arefs{n} }
\item \Idef{yamagata}{Yamagata University, Yamagata, 992-8510 Japan\Arefs{k} }
\end{Authlist}
%
%
\vspace*{-\baselineskip}\renewcommand\theenumi{\alph{enumi}}
\begin{Authlist}
\item \Adef{a}{Also at Instituto Superior T\'ecnico, Universidade de Lisboa, Lisbon, Portugal}
\item \Adef{aa}{Also at Department of Physics, Pusan National University, Busan 609-735, Republic of Korea and at Physics Department, Brookhaven National Laboratory, Upton, NY 11973, U.S.A. }
\item \Adef{bb}{Supported by the DFG Research Training Group Programme 1102  ``Physics at Hadron Accelerators''}
\item \Adef{b}{Also at Chubu University, Kasugai, Aichi, 487-8501 Japan\Arefs{k}}
\item \Adef{c}{Also at KEK, 1-1 Oho, Tsukuba, Ibaraki, 305-0801 Japan}
\item \Adef{c1c}{Present address: Universit\"at Bonn, Helmholtz-Institut f\"ur Strahlen- und Kernphysik, 53115 Bonn, Germany}
\item \Adef{cc}{Also at Moscow Institute of Physics and Technology, Moscow Region, 141700, Russia}
\item \Adef{x}{present address: RWTH Aachen University, III. Physikalisches Institut, 52056 Aachen, Germany}
\item \Adef{f}{Supported by the German Bundesministerium f\"ur Bildung und Forschung}
\item \Adef{g}{Supported by Czech Republic MEYS Grants ME492 and LA242}
\item \Adef{h}{Supported by SAIL (CSR), Govt.\ of India}
\item \Adef{i}{Supported by CERN-RFBR Grants 08-02-91009 and 12-02-91500}
\item \Adef{j}{\raggedright Supported by the Portuguese FCT - Funda\c{c}\~{a}o para a Ci\^{e}ncia e Tecnologia, COMPETE and QREN, Grants CERN/FP/109323/2009, CERN/FP/116376/2010 and CERN/FP/123600/2011}
\item \Adef{k}{Supported by the MEXT and the JSPS under the Grants No.18002006, No.20540299 and No.18540281; Daiko Foundation and Yamada Foundation}
\item \Adef{l}{Supported by the DFG cluster of excellence `Origin and Structure of the Universe' (www.universe-cluster.de)}
\item \Adef{ll}{Supported by EU FP7 (HadronPhysics3, Grant Agreement number 283286)}
\item \Adef{m}{Supported by the Israel Science Foundation, founded by the Israel Academy of Sciences and Humanities}
\item \Adef{n}{Supported by the Polish NCN Grant DEC-2011/01/M/ST2/02350}
\item [{\makebox[2mm][l]{\textsuperscript{*}}}] Deceased
\end{Authlist}

\newpage

\section{Introduction}
The description of the nucleon spin structure is still one of the open issues in
hadron physics. In the last decades major progress in this field has been made
by an interplay between new experimental results and the development of
non-collinear QCD. The first information on transverse spin and transverse
momentum effects become available recently. Presently, the complete description of
quarks in the nucleon includes all possible correlations between quark spin,
quark transverse momentum and nucleon
spin~\cite{Kotzinian:1994dv,Mulders:1995dh}. At leading twist, these
correlations are described for each quark flavour by eight transverse momentum
dependent (TMD) parton distribution functions (PDFs). After integration over
transverse momentum, only three of them survive, namely the number density, the
helicity and the transversity PDFs. One way to access experimentally these TMD
PDFs is via semi-inclusive lepton--nucleon deep inelastic scattering (SIDIS),
i.e. by studying deep-inelastic scattering (DIS) with detection of at least one
of the produced hadrons. When the target nucleon is transversely polarised, the
SIDIS cross section exhibits different azimuthal
modulations~\cite{Bacchetta:2006tn} in different combinations of the two angles
$\phi_S$ and $\phi_h$. These are the azimuthal angles of the initial nucleon
transverse spin vector and of the produced hadron momentum in a 
reference system, in which the $z$-axis is the virtual photon direction and the
$x-z$ plane is the lepton plane according to Ref.~\cite{Bacchetta:2004jz}. The
amplitudes of the modulations in the cross section (the so-called transverse
spin asymmetries) are proportional to convolutions of TMD PDFs with TMD
fragmentation functions. The two most thoroughly studied transverse spin
asymmetries are the Collins and Sivers asymmetries. The Collins asymmetries
allow access to the transversity PDFs coupled to the Collins fragmentation
functions~\cite{Collins}. The Sivers asymmetries give access to the Sivers
PDFs~\cite{Sivers}, which describe the correlations between quark transverse
momentum and nucleon spin. A Sivers PDF always appears in combination with an
`ordinary' (unpolarised) fragmentation function that describes the fragmentation
of a quark into a hadron.

In this Paper, we present results on the Collins and Sivers asymmetries for pions
and kaons produced on transversely polarised protons in a NH$_3$ target. These
measurements are in line with the set of measurements done by the COMPASS
Collaboration in the last years. Results on polarised deuterons were obtained
for unidentified hadrons~\cite{ref:transvdeut2}, pions and
kaons~\cite{ref:transvdeut3}, and on polarised protons for charged unidentified
hadrons~\cite{Alekseev:2010rw,ref:transvp,ref:transvpsiv}. The results presented
in this Letter are extracted from all available COMPASS data taken in 2007 and
2010 using transversely polarised protons. For the measurements in 2007 and 2010,
a similar spectrometer configuration was used. As compared to the measurements
on transversely polarised deuterons, the measurements on transversely polarised
protons benefit from a major upgrade of the apparatus performed in 2005. Of
particular relevance for these measurements is the upgrade of the RICH
detector~\cite{ref:rich}, which led to improved efficiency and purity for the
samples of identified particles, and the use of a new target solenoid magnet
with a polar angle acceptance of 180 mrad as compared to the 70 mrad of the
magnet used until 2005. Measurements of these asymmetries by the HERMES experiment
exist~\cite{ref:herm, ref:hermsiv} in a different kinematic range. Comparison
with these results are also presented in the Paper.

\section{Apparatus and data selection}
\label{selection}
The COMPASS spectrometer~\cite{Abbon:2007pq} is in operation in the North Area
of CERN since 2002. The $\mu^+$ beam provided by the M2 beam line had a momentum
of $160\ \mrf{GeV}/c$, a momentum spread $\Delta p/p = \pm 5\%$, and a longitudinal
polarisation of $-80\%$ that originated from the $\pi$-decay mechanism. The mean
beam intensity was about $2.3\times10^7 \mu^+/s$ and $4\times10^7 \mu^+/s$ with
spill lengths of 4.8~s and 10~s in 2007 and 2010, respectively.

The target consisted of three cylindrical cells of 4~cm diameter, each
separated by gaps of 5~cm. The length of the central cell was 60~cm and that
of the two outer ones 30~cm. For the measurement of transverse spin effects, the
target material was polarised along the vertical direction. In order to reduce
systematic effects, neighbouring cells were polarised in opposite directions,
which allows for simultaneous data taking with both target spin directions. To
further minimise systematic effects, the polarisation of each cell was reversed
every 4--5 days. During the 2007 data taking, a total amount of $12\times10^9$
events (440~TB) was recorded in six periods, each consisting of two sub-periods
of data taking with opposite polarisation. In 2010 about $37\times10^9$ events
(1.9~PB) were recorded over twelve periods.

Only events with a photon virtuality $Q^2 >1\ (\mrf{GeV}/c)^2$ and a mass of the
hadronic final state $W>5\ \mrf{GeV}/c^2$ have been used to ensure the
kinematic region of DIS. The upper limit on the fractional energy of the virtual
photon ($y$) was set to 0.9 to reduce uncertainties due to electromagnetic radiative
corrections and contamination from pion decay. A lower limit on $y$ is
required to ensure a good resolution in this variable. In the standard analysis
this limit has been set to 0.1. A complementary sample with $0.05<y<0.1$ was 
also studied, mainly to address the $Q^2$ dependence of the asymmetries.
The Bjorken variable $x$ covers the range from 0.003 to 0.7. A minimum value of
$0.1\ \mrf{GeV}/c$ for the hadron transverse momentum $p_T^h$ with respect to the
virtual photon direction was required to ensure good resolution in the
measured azimuthal angle. A minimum value for the relative hadron energy $z$
with respect to the virtual photon energy is needed to select hadrons from the
current fragmentation region. This value has been set to 0.2 for the standard
sample, while a complementary lower-$z$ region $(0.1<z<0.2)$ was also studied.

The stability of the apparatus during data taking is crucial. Therefore, various
tests were performed using the 2007 and 2010 data, as described
in~\cite{Alekseev:2010rw,ref:transvp,ref:transvpsiv}. As a result from these
quality tests, only four periods of data taking in 2007 were used for the
analysis of the Sivers asymmetries, while for the Collins asymmetry all six
periods were used. This can be understood as the Sivers asymmetry is very
sensitive to instabilities in the spectrometer acceptance, because it represents
the amplitude of the modulation that depends on the azimuthal angle of the
hadron transverse momentum with respect to the target spin vector, which is
aligned along a fixed direction. Due to improved detector stability, all periods
of 2010 could be used for the extraction of both asymmetries.

\section{Particle identification}
\subsection{Charged pions and kaons}
The RICH detector information was used to identify charged hadrons as kaons
and pions. The pattern of the detected photons in the detector was analysed
taking into account the predicted path of the charged particle to compute
likelihood values~\cite{ref:richsoftware} for each reconstructed track entering
the RICH acceptance. The likelihoods $\mathcal{L}$ were computed for different
mass hypotheses ($\mathcal{L}_{M}$, with $M= \pi$, $K$, $p$, $e$) and for the
hypothesis of absence of signal, namely the so called background hypothesis
($\mathcal{L}_{back}$). A mass value is attributed to a track if the likelihood
for the corresponding mass hypothesis is the largest. In addition, cuts on the
ratio of the largest likelihood value to the second largest one were added to
improve the figure of merit given by the product of the identification
efficiency and sample purity, as defined below.  Pions and kaons were identified
in the momentum range between the Cherenkov threshold (about
$2.6\ \mrf{GeV}/c$ for pions, $9\ \mrf{GeV}/c$ for kaons) and
$50\ \mrf{GeV}/c$. A specific cut on the ratio
$\mathcal{L}_{K}/\mathcal{L}_{back}$ was applied to minimise the contamination
of protons in the kaon sample in the momentum range between the kaon threshold
and the proton one (about $18\ \mrf{GeV}/c$). In Fig.~\ref{fig:mom} the
distributions in momentum ($p$), $z$ and $Q^2$  for identified pions and kaons are
shown and compared to those for unidentified hadrons. The mean values of the
$p_T^h$, $z$ and $Q^2$ distributions as a function of $x$ are shown in
Fig.~\ref{fig:meanv}.

Both the sample purity and the identification efficiency were measured and
found to be compatible in the two data taking years as well as in the
different periods of each year. The particle identification efficiencies and
misidentification probabilities were determined using samples of pions
from the $K^0$ decay and of kaons from the $\phi$ decay. The
efficiencies are about 97\% for pions and 94\% for kaons. These values start
to decrease for momenta about $30\ \mrf{GeV}/c$ and reach values of 60\% in the
larger momentum region. In order to achieve high values of the sample purity,
the misidentification probabilities between pions and kaons were kept as
low as a few percent even at the largest momentum values. The purity is
defined as the fraction of $K$ ($\pi$) inside the identified $K$ ($\pi$)
sample and depends also on the different population of the various particle
types. It was evaluated from the particle identification efficiencies and
misidentification probabilities and the number of identified kaons and pions,
since the proton contribution is very small as already mentioned. The average
purity values for pions are above 99\%. The kaon purity is shown in
Fig.~\ref{fig:purity} as a function of $x$, $z$, and $p_T^h$; it is about 94\%
with a mostly mild dependence on the variables. The strongest dependence is
visible in the large $z$ region for the negative kaon sample, which is due to
the increasing ratio of pions to kaons. The resulting statistics for charged pions
and kaons after all cuts are shown in Table~\ref{tab:finalstat}.

\begin{table}[ht!] 
  \begin{center}
  \caption[2010 final statistics]{Final statistics for 2007 and 2010 for
   identified charged pions and kaons and neutral kaons.}
	   \label{tab:finalstat}
   	   \begin{tabular}{lrrrrr}
      \hline
           Year &\multicolumn{5}{c}{Number of particles ($\times 10^{-6}$)}\\
    	   	& $\pi^+$ & $\pi^-$ &$K^+$ &$K^-$ &$K^0$\\
	  \hline
	   2007 (Collins) & 10.77  &	9.41  	&1.79  	&1.10  &	0.37  \\
   		 2007 (Sivers) &6.84  &	5.97  	&1.12  	&0.69  & 0.25  \\
   	  	2010 & 27.26  	&23.72  	&4.48  	&2.71  & 1.00  \\	
 	  \hline
  \end{tabular}
   \end{center}
  \end{table}
\bef
\bec
\includegraphics[width=0.329\textwidth]{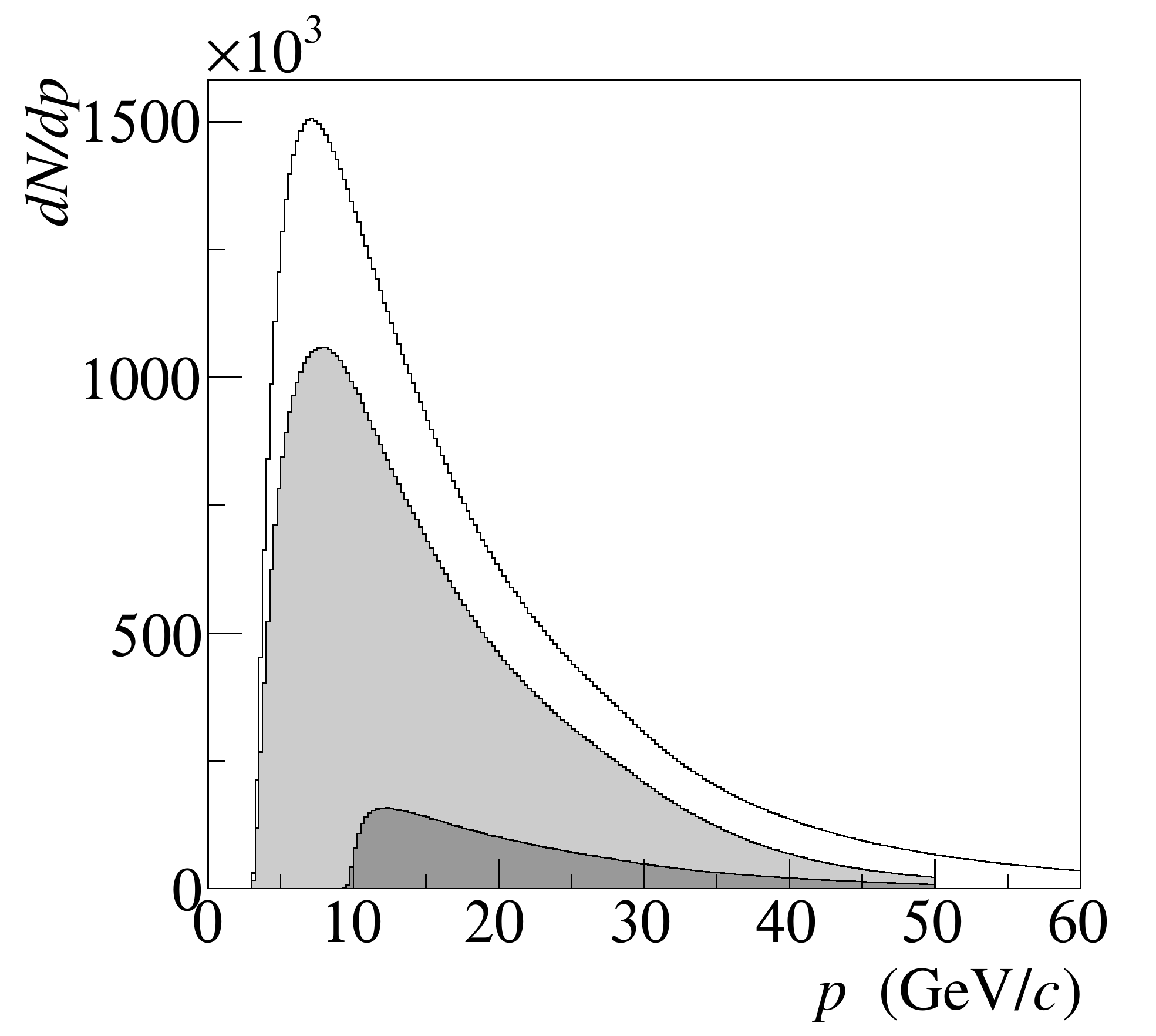}
\includegraphics[width=0.329\textwidth]{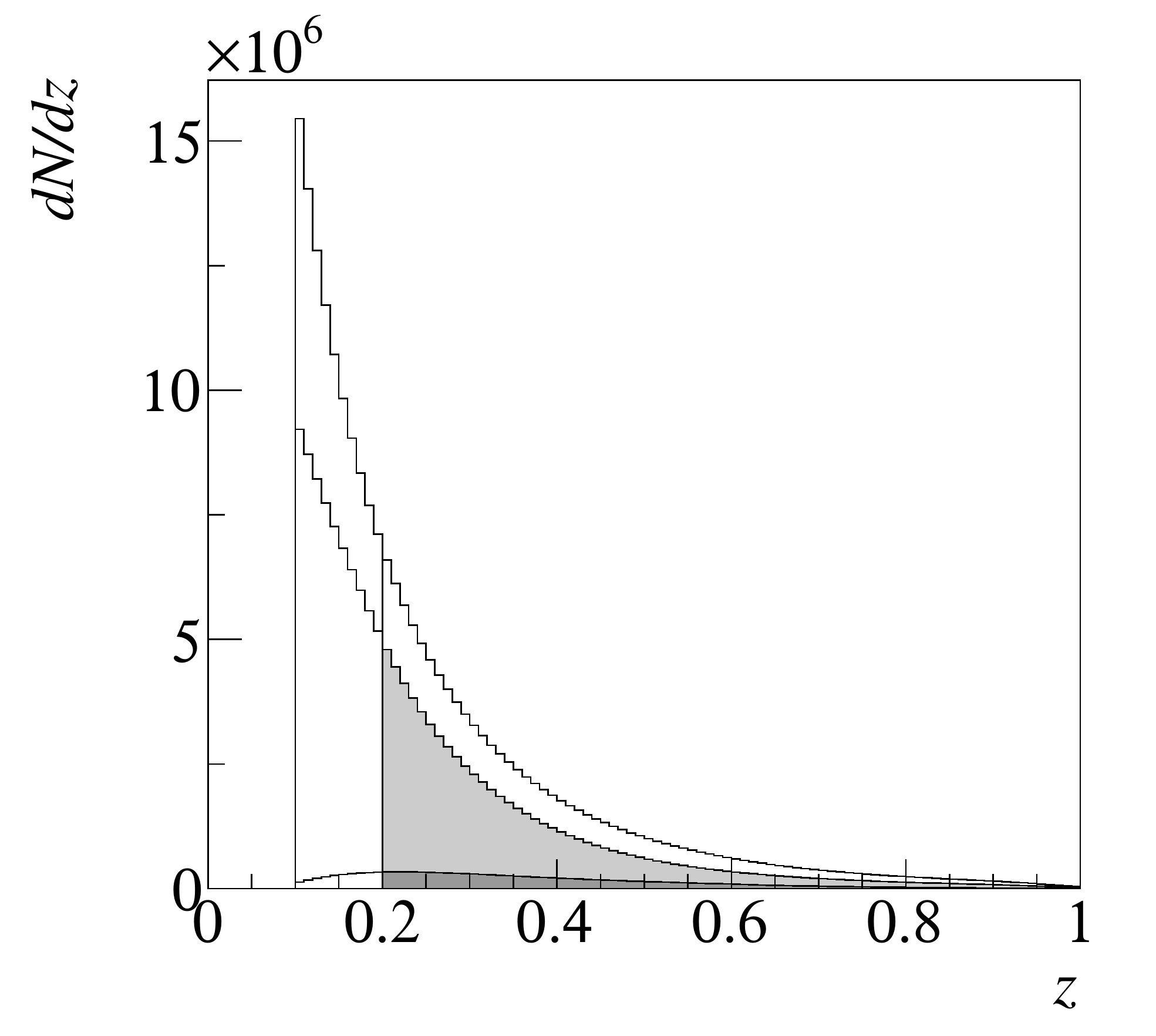}
\includegraphics[width=0.329\textwidth]{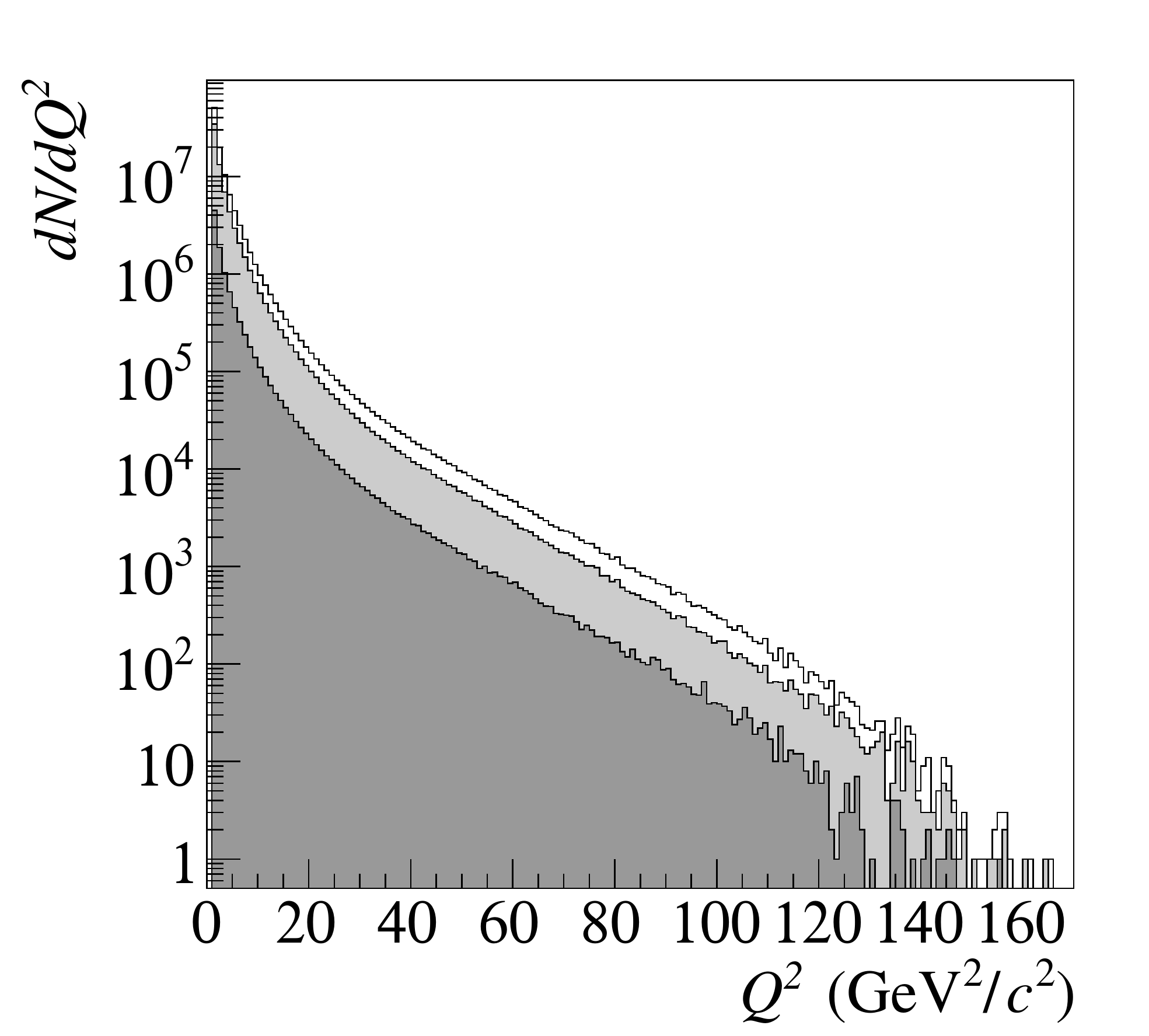}
\caption{Momentum ($p$) (left), relative energy $z$ (centre), $Q^2$ (right)
 distribution of the unidentified hadrons (white), pions (light grey) and kaons
 (dark grey).\label{fig:mom}}
\eec
\eef
\bef
\bec
\includegraphics[width=0.329\textwidth]{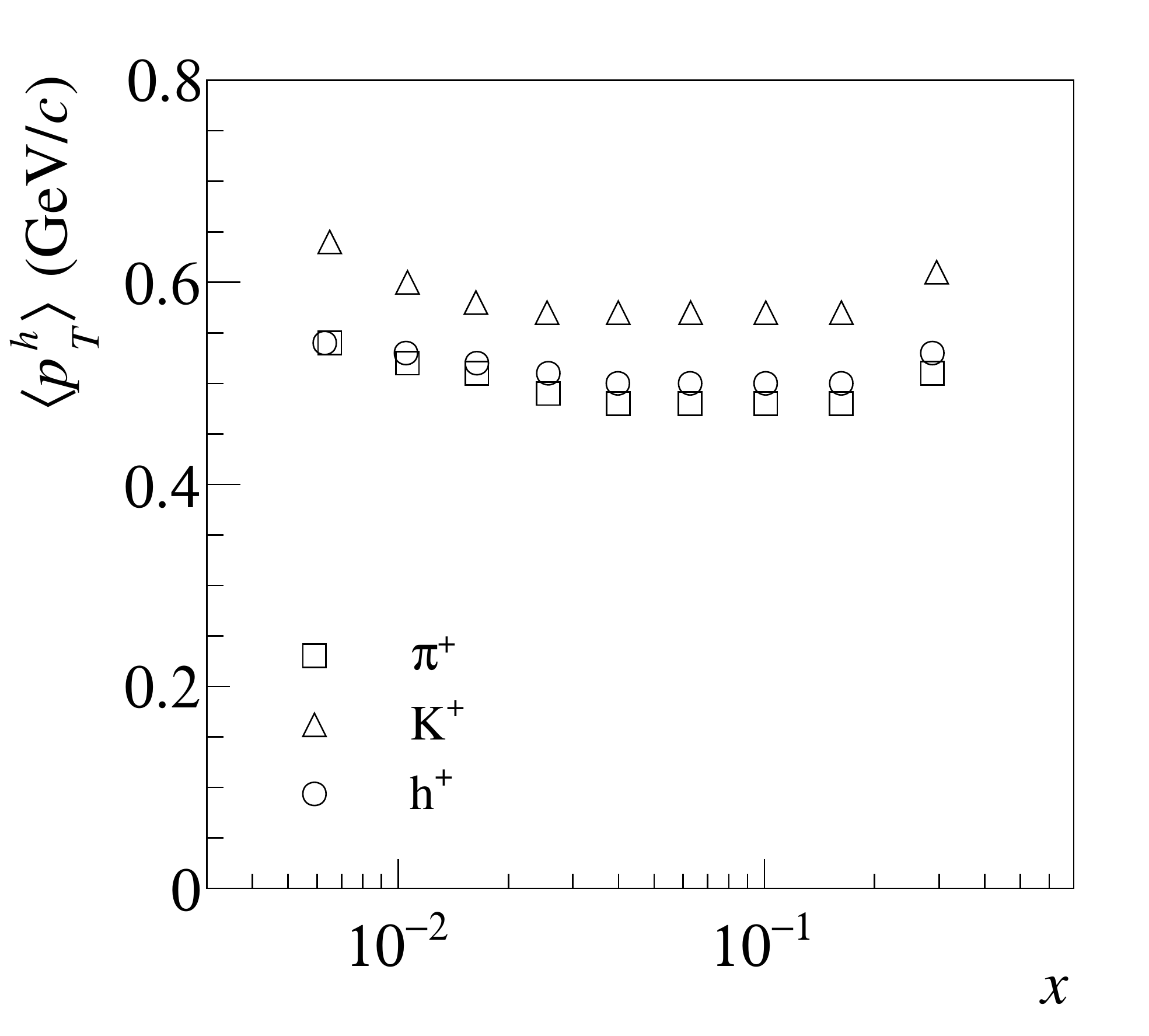}
\includegraphics[width=0.329\textwidth]{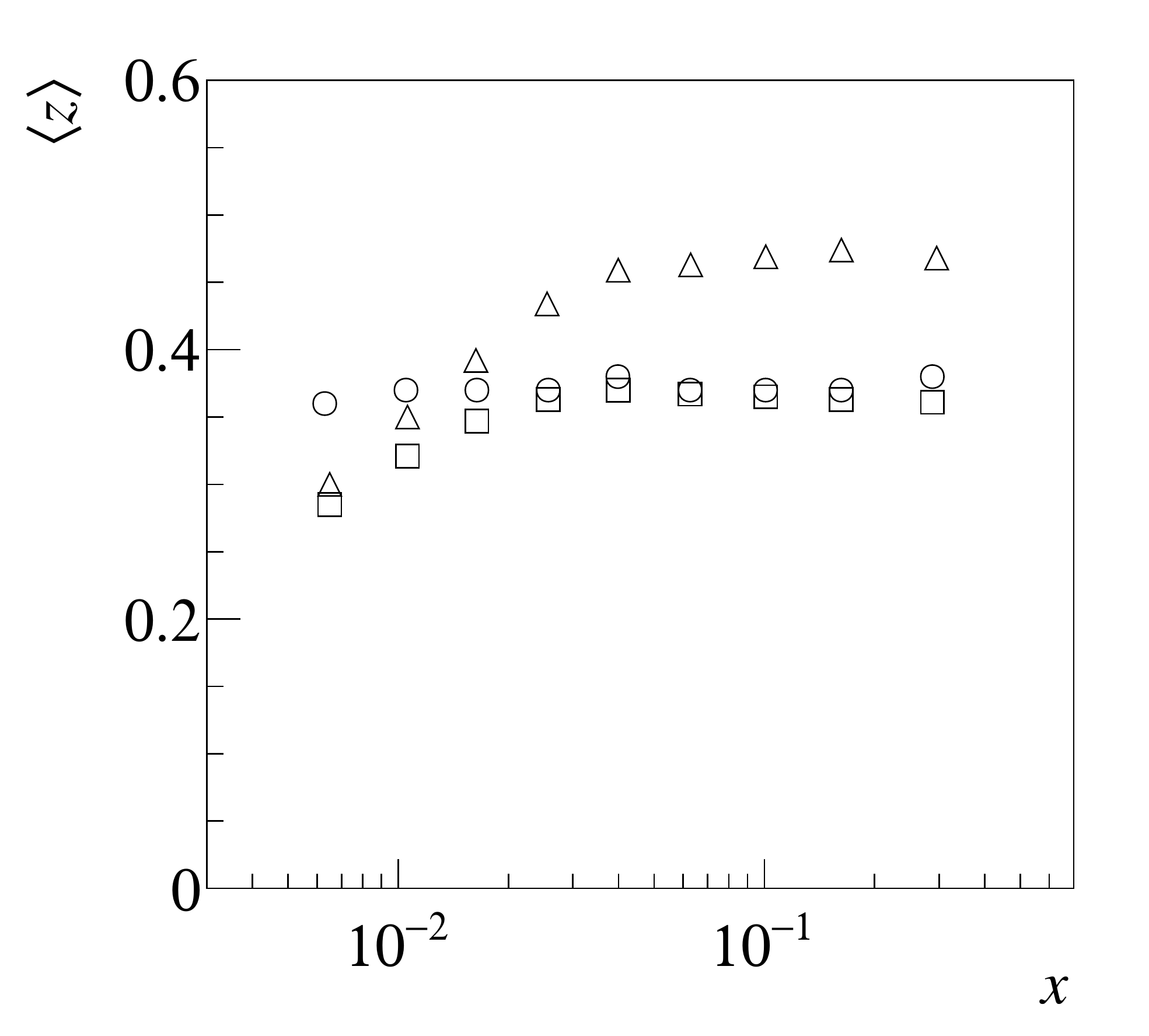}
\includegraphics[width=0.329\textwidth]{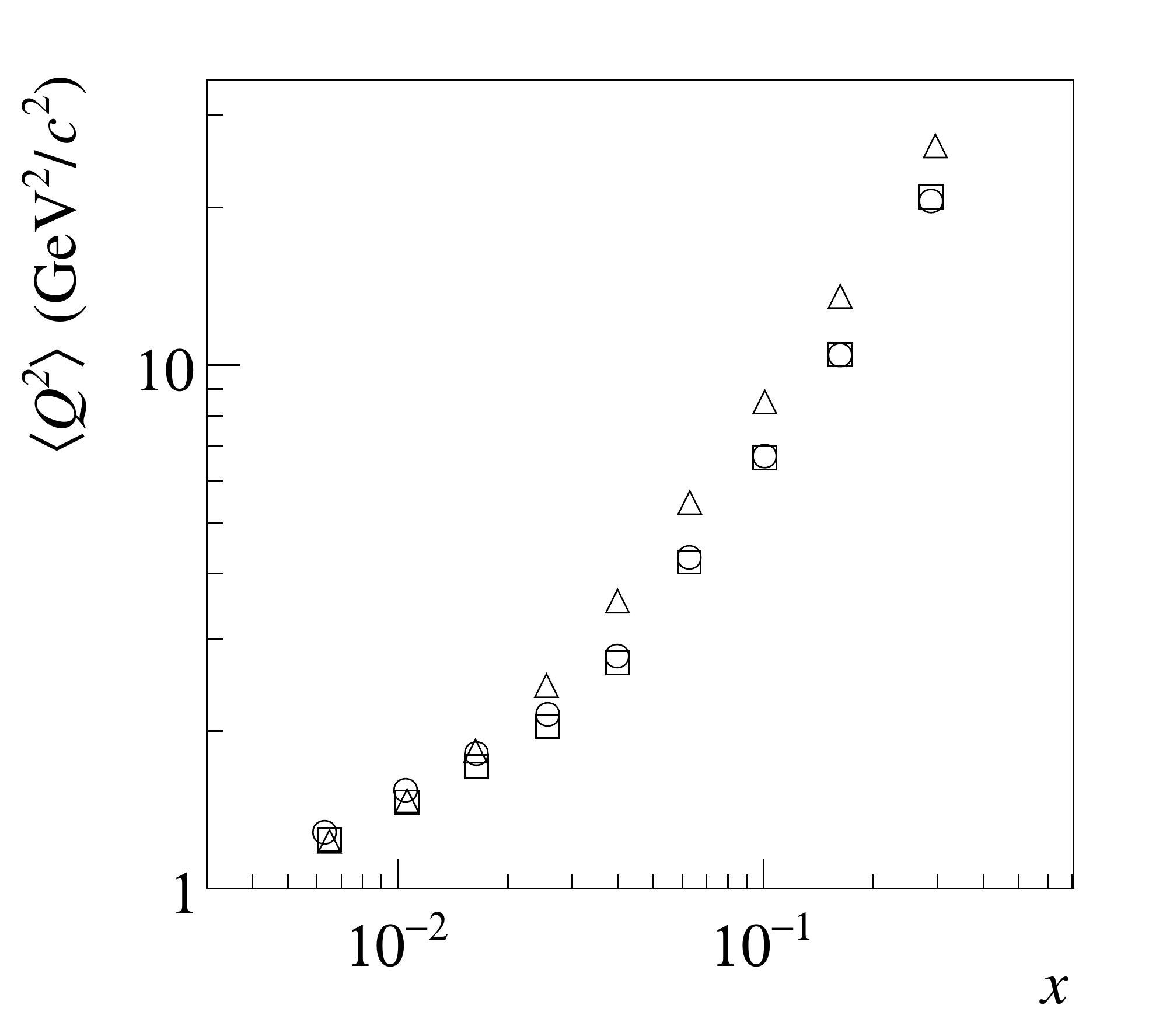}
\caption{Mean values of the transverse momentum $p_T^h$ (left), relative energy
 $z$ (centre), $Q^2$ (right) of the unidentified hadrons (circles), pions
 (squares) and kaons (triangles).\label{fig:meanv}}
\eec
\eef
\bef
\bec
\includegraphics[width=0.8\textwidth]{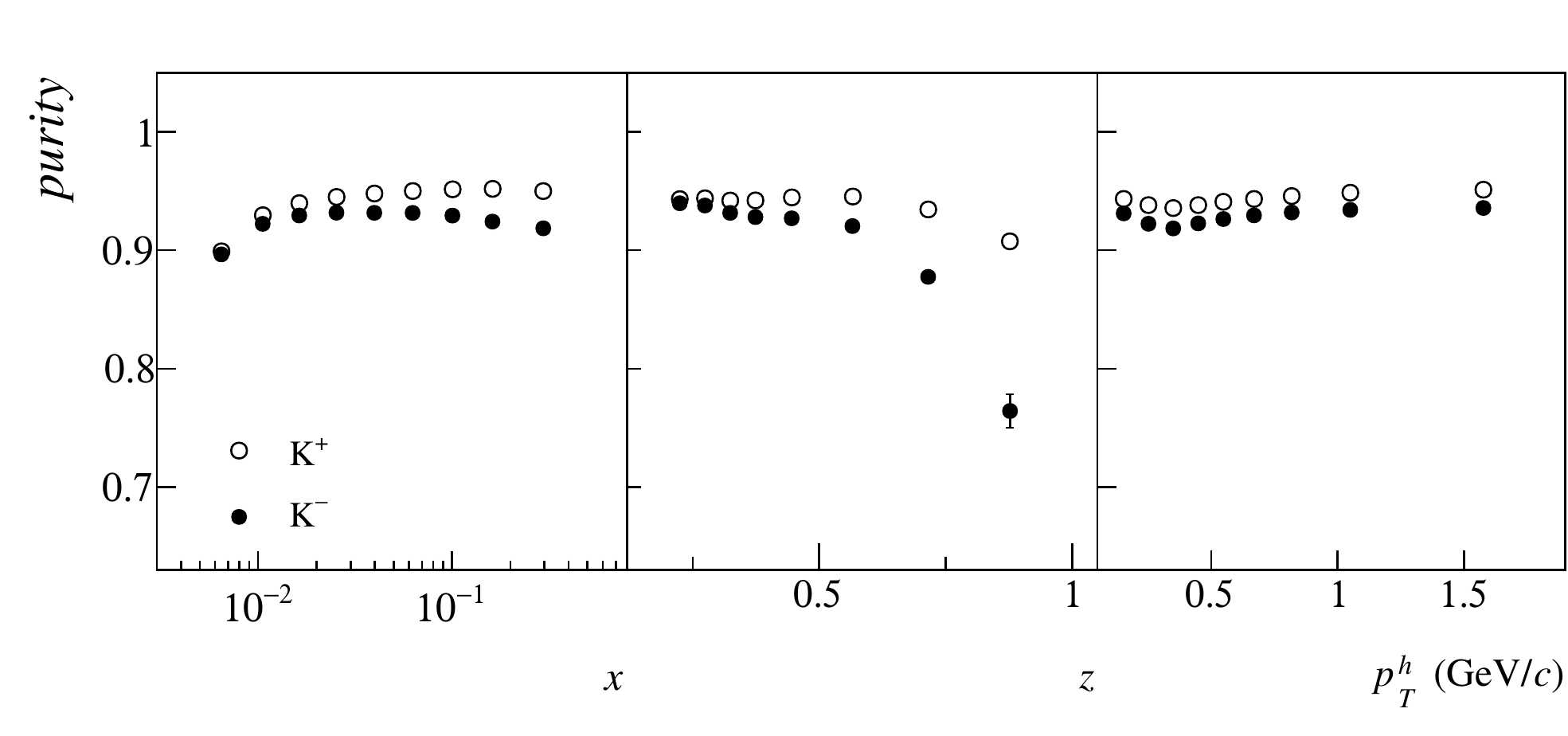}
\caption{ Purity of the identified positive and negative kaons as a function of
 $x$, $z$, $p_T^h$.\label{fig:purity}}
\eec
\eef

\subsection{$K^0$ identification}
The $K^0$ identification is based on the detection of two oppositely charged
tracks coming from a secondary vertex, for which the 2-pion invariant mass lies
in the window $m_{K^0}\pm 20\ \mrf{MeV}/c^2$. A separation between the primary
and the secondary vertex of 
at least 10~cm was required. Furthermore, the angle between the reconstructed
momentum vector of the track pair and the vector connecting the primary and
secondary vertices was required to be smaller than 10~mrad. On the left side of
Fig.~\ref{fig:armmass}, the Armenteros-Podolansky plot of the hadron pair is
shown, in which the transverse momentum $p_T$ of one hadron with respect to the
sum of hadron momenta is shown as a function of the difference of the
longitudinal momenta over their sum, $(p_{L1}-p_{L2})/(p_{L1}+p_{L2})$. The
$K^0$ band is clearly visible as well as the $\Lambda$ and $\bar{\Lambda}$
bands. In order to exclude the contamination by the $\Lambda/\bar{\Lambda}$
signal, the $p_T$ region between $80\ \mrf{MeV}/c$ and $110\ \mrf{MeV}/c$ was
excluded. The background from $e^+e^-$ pairs was suppressed by a lower cut on
$p_T$ at $40\ \mrf{MeV}/c$.  For the detected $K^0$s, the difference between
their mass value and the PDG~\cite{pdg2012} value is shown in the right panel of
Fig.~\ref{fig:armmass}, where the vertical lines at $\pm 20\ \mrf{MeV}/c^2$
enclose the $K^0$s used for further analysis.  The same cuts on $z$ and $p_T$ as
for the charged hadron samples were applied to the neutral kaons. The resulting
statistics for $K^0$ are given in Table~\ref{tab:finalstat}. 
\begin{figure}[!ht]
\centering
\includegraphics[width=0.497\linewidth]{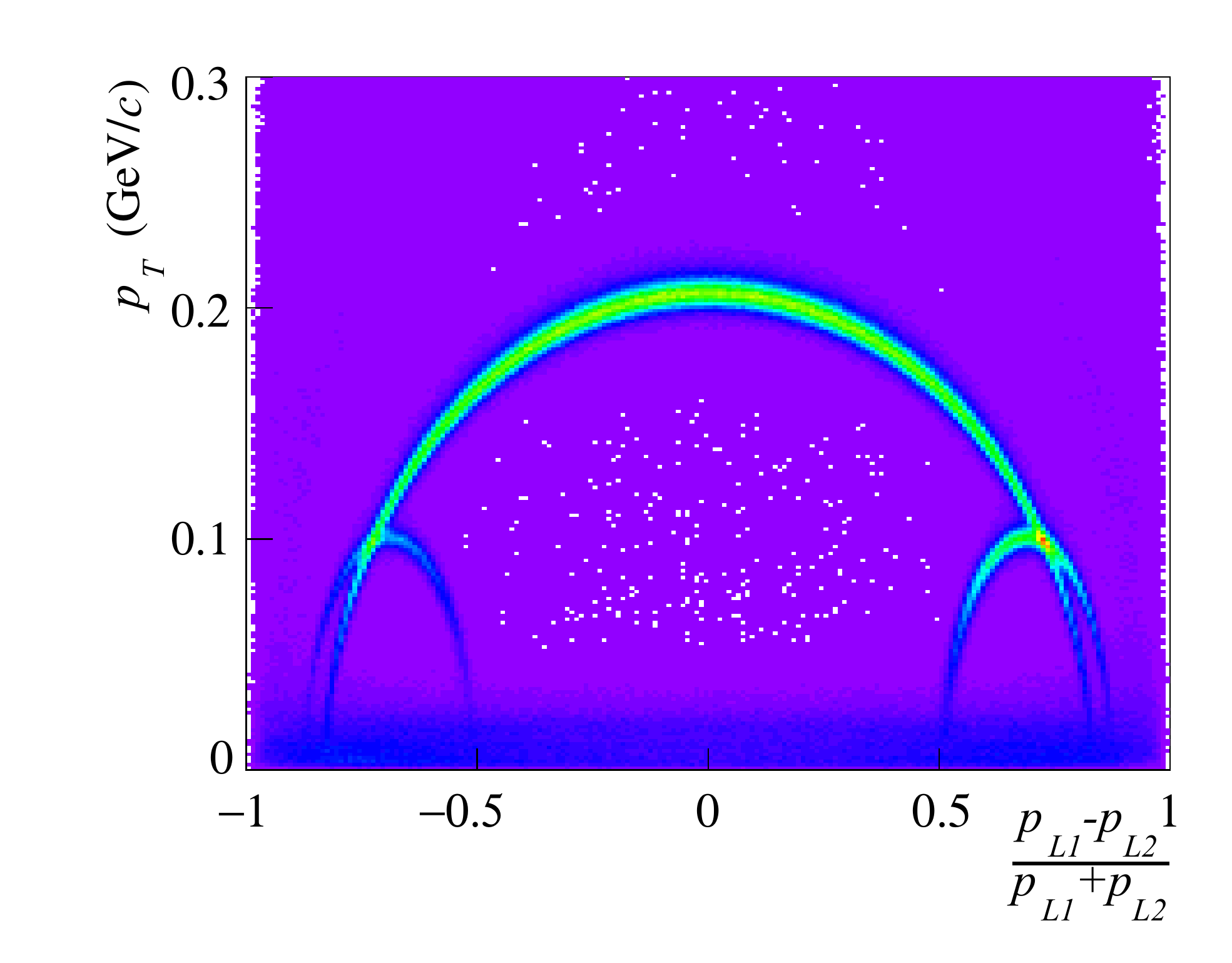}
\includegraphics[width=0.497\linewidth]{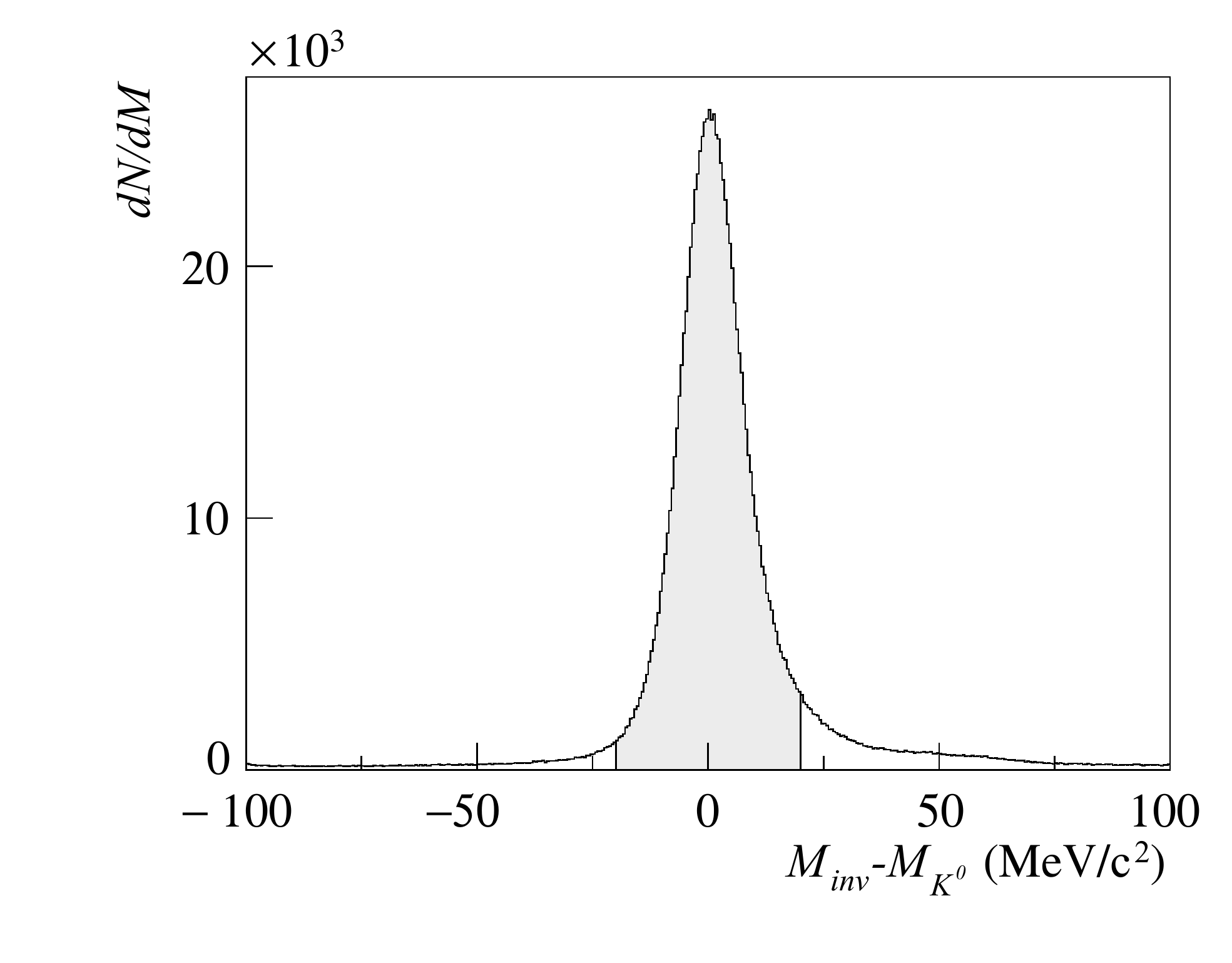}
\caption[Left: Armenteros-Podolansky plot of the hadron pair. Right: Difference
  of the invariant mass of the hadron pair.]{Left: Armenteros-Podolansky plot of
  the hadron pair. Right: Difference of the invariant mass of the hadron pair
  and the PDG value of the $K^0$ mass. The mass range used for the analysis is
  shaded.}
\label{fig:armmass}
\end{figure}

\section{Results}
In order to extract the transverse spin asymmetries from the data, the same
procedure as described in Refs.~\cite{ref:transvp,ref:transvpsiv} was used.
The asymmetries were evaluated in bins of the kinematic variables $x$, $z$,
or $p_T^h$, using the same binning as in our previous
analyses~\cite{ref:transvdeut3}. All the numerical results are available on
HEPDATA. The ($\phi_S$,$\phi_h$) distributions from the
different target cells and sub-periods were fitted using an extended maximum
likelihood estimator~\cite{Alekseev:2010rw}, and the eight transverse spin
asymmetries expected in the SIDIS process were extracted simultaneously.

The resulting $\sin(\phi_h+\phi_s)$ and  $\sin(\phi_h-\phi_s)$ modulations
yield the Collins and Sivers asymmetries, respectively, after
division by {\em i}) the target material dilution factor, {\em ii}) the
average target proton polarisation, and for the Collins asymmetry {\em iii})
the transverse spin transfer coefficient.
The dilution factor of the ammonia target, taking also into account the
electromagnetic radiative corrections, was evaluated in each $x$
bin~\cite{ref:transvp}: it increases with $x$ from 0.14 to 0.17. As a function
of $z$ and $p^h_T$ the dilution factor turns out to be almost constant with an
average value of 0.15.

A non-flat azimuthal acceptance introduces correlations between the various
modulations resulting from the fit. The correlation coefficients for Collins
vs. Sivers asymmetries are found to be small and below 0.2 for all
bins. Moreover, the asymmetries measured along different projections of the ($x$,
$z$, $p_T^h$) phase space are statistically correlated, because the overall
sample of events is the same. In the case of COMPASS, these correlation coefficients
for the Collins and for the Sivers asymmetries are all smaller than about 0.3,
but non-negligible, so that they should be taken into account in any global fit.
They are slightly different for kaons and pions due to the different kinematic
coverage of the two samples.
 
In order to estimate the systematic uncertainties, several tests were performed
based on our previous work for the charged
hadrons~\cite{Alekseev:2010rw,ref:transvp,ref:transvpsiv}. The effect of changes
in the azimuthal acceptance between the data sets used to extract the
asymmetries was quantified building false asymmetries, namely assuming a wrong
polarisation direction in the target cells. The Collins and Sivers asymmetries
were extracted splitting the data according to the scattered muon direction (up
and down, left and right), and the statistical compatibility of the results was
checked.  No such false asymmetry was observed within the accuracy of the
measurement and the point-to-point systematic uncertainties were evaluated from
these tests as a fraction of the statistical error. For 2010, this fraction is
0.6 and for 2007 it ranges between 0.5 and 0.7.  The systematic uncertainty due
to particle misidentification is very small and included in these fractions.
All results are subject to a 3\% scale uncertainty that results from the
uncertainties in the target polarisation and dilution factor.

For both years of data taking, the asymmetries were evaluated in each period and
their compatibility was checked. While for 2010 this test shows good agreement
among the results of the different periods, for 2007 it introduces an additional
source of systematic uncertainties for the Sivers asymmetries. Very much like in
the case of unidentified hadrons, an additional absolute uncertainty of $\pm 0.012$ was
assigned to the $\pi^+$ Sivers asymmetry. This value is taken to be half of
the difference between the mean asymmetries evaluated using the data from the
beginning and the end of the 2007 data taking. Figure~\ref{fig-1} shows the
Collins and Sivers asymmetries for pions as a function of $x$ from the two data
taking years, obtained as weighted mean of the asymmetries from the different
periods. The two measurements are in good agreement. The substantial improvement
of the statistical precision of the 2010 data with respect to the 2007 data
amounts to a factor of $1.6$ for the Collins asymmetry and of $1.9$ for the
Sivers asymmetry.
\begin{figure}
\centering 
\includegraphics[width=0.497\textwidth]{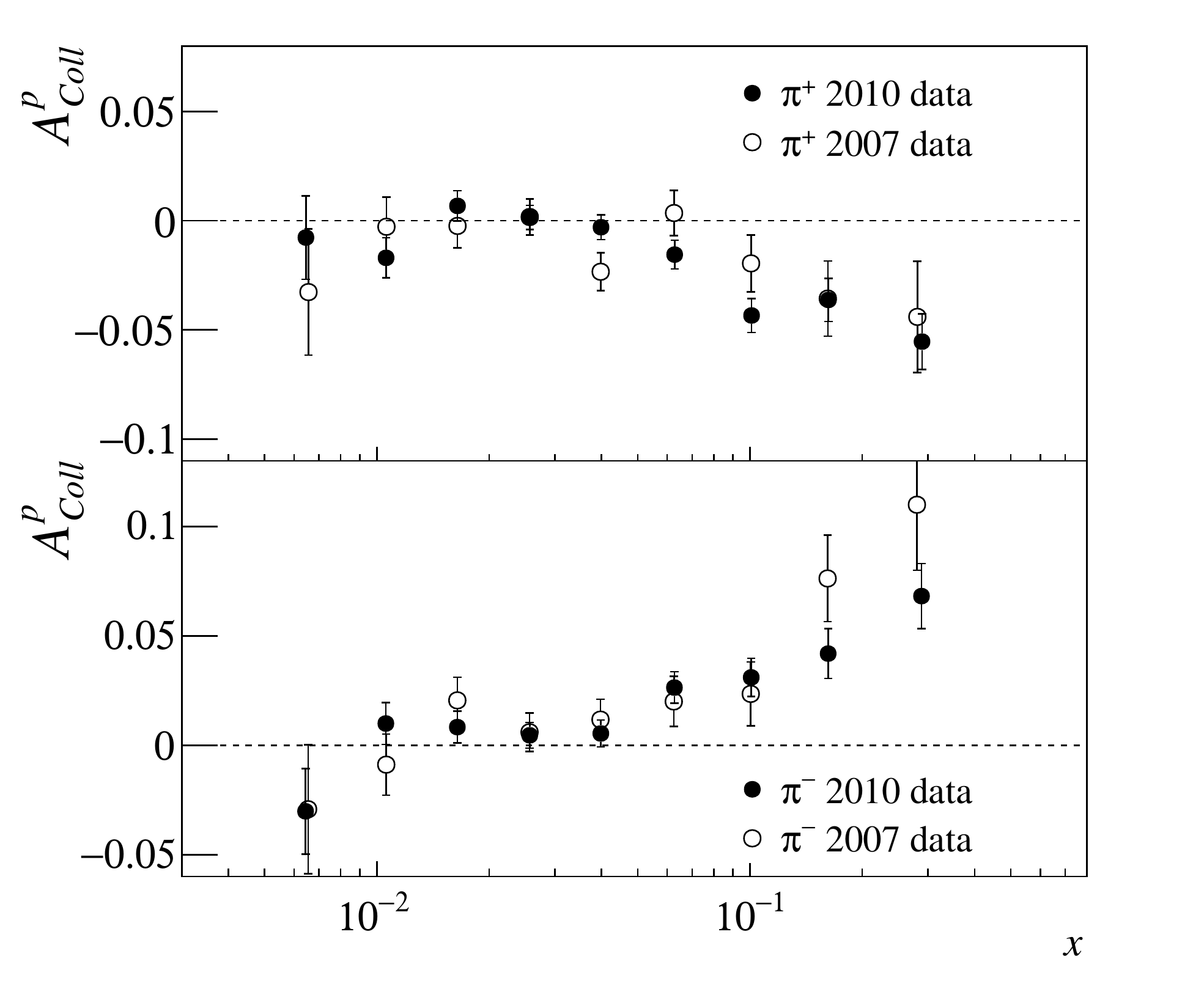}
\includegraphics[width=0.497\textwidth]{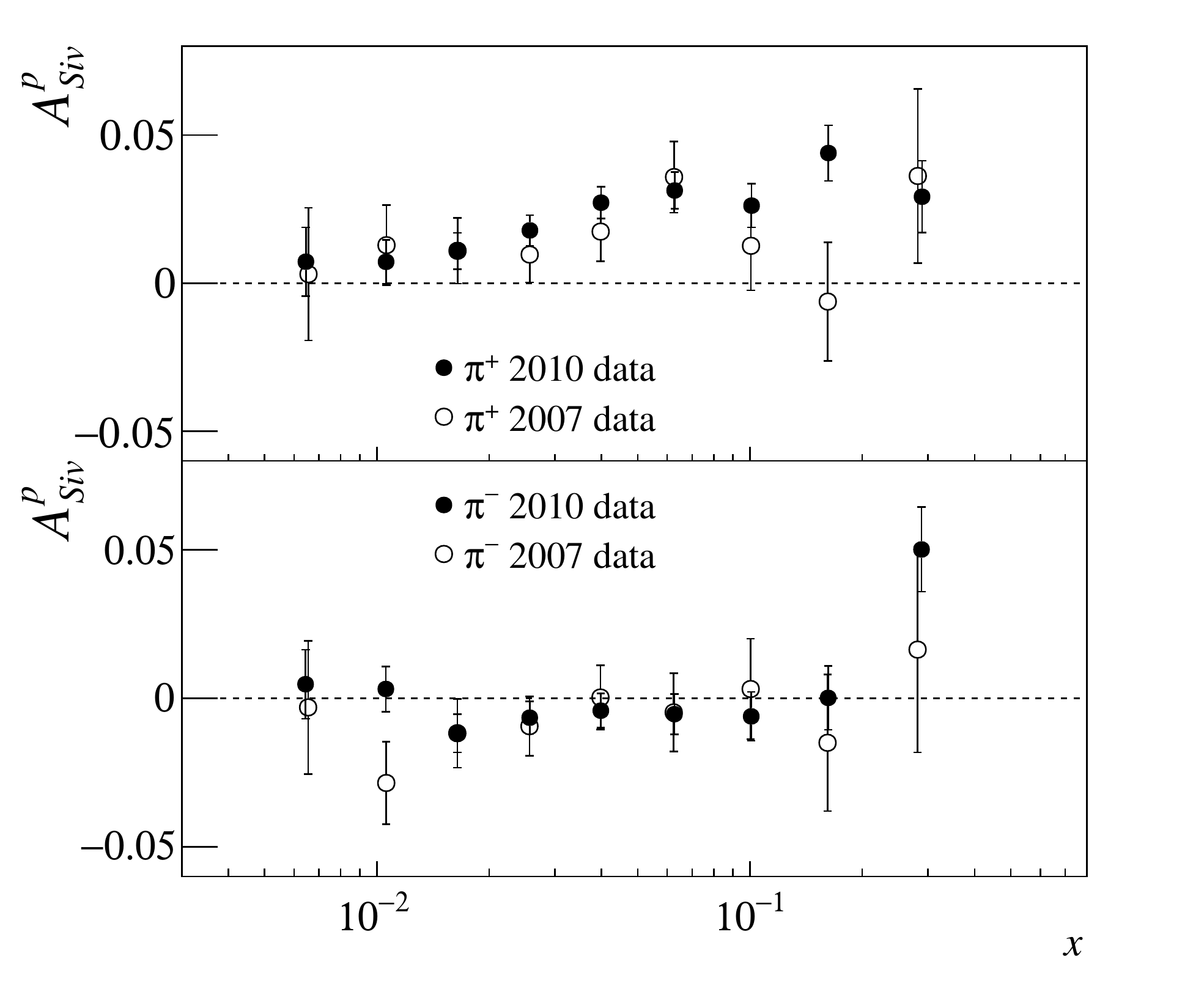}
\caption{Left: comparison between the Collins asymmetries for pions as a
 function of $x$, extracted from 2007 and 2010 data taking. Right: the same
 comparison for the Sivers asymmetries.\label{fig-1}}
\end{figure}
The final results were obtained combining the two samples, taking into
account the different statistical and systematic uncertainties. The resulting
systematic uncertainties are about 0.6 of the statistical ones for all the
particle types.

The Collins asymmetries as a function of $x$, $z$, or $p_T^h$ measured by COMPASS
for pions and kaons on transversely polarised protons are shown in
Fig.~\ref{fig-call}. The pion asymmetries are very similar to the unidentified
hadron asymmetries~\cite{ref:transvp}: at small $x$ they are compatible with
zero, while in the valence region they show an increasing signal, which has
opposite sign for $\pi^+$ and $\pi^-$. This naively indicates that the
unfavoured and favoured Collins fragmentation functions have opposite sign.
\begin{figure}
\centering \includegraphics[width=0.9\textwidth]{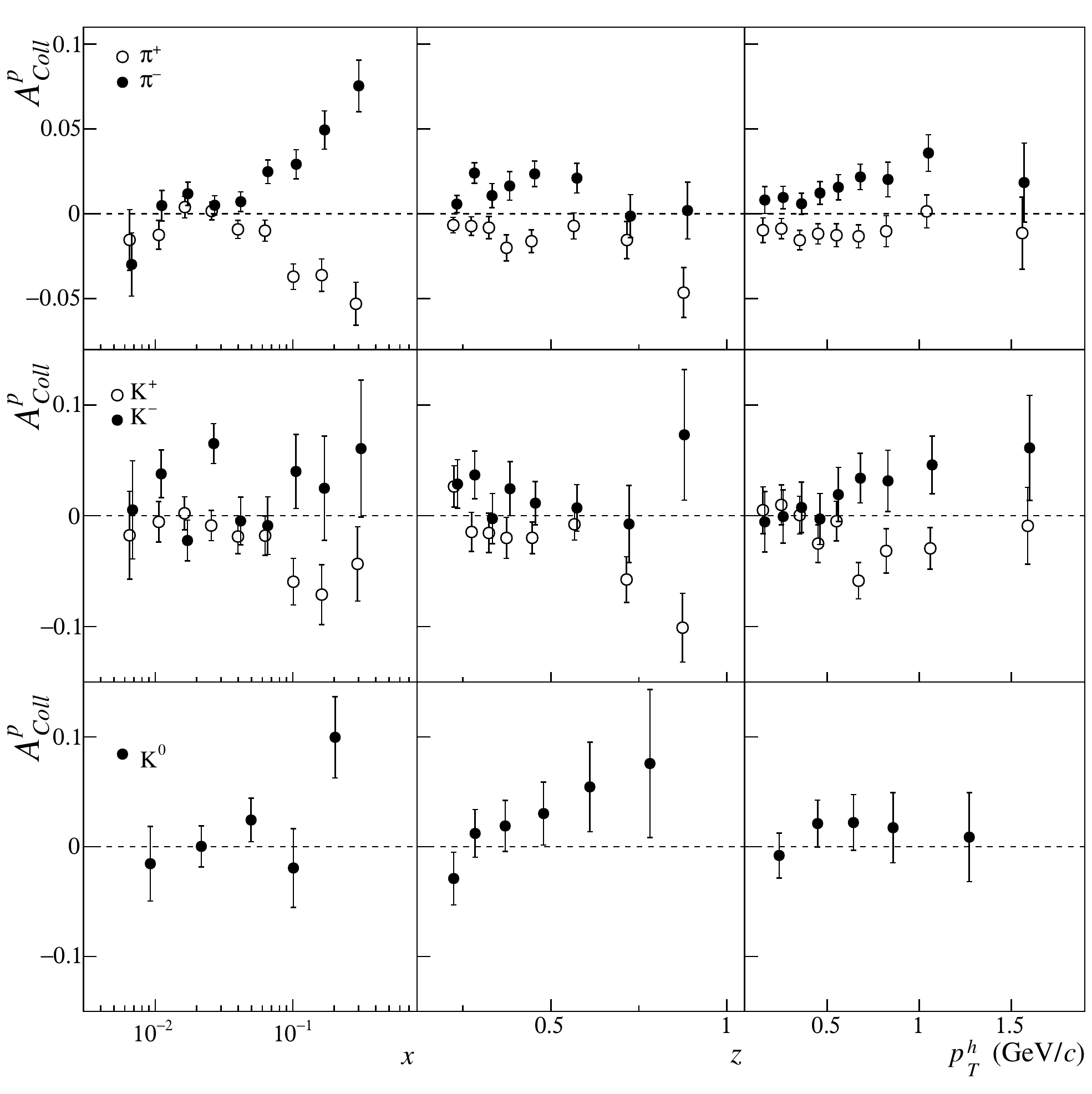}
\caption{The Collins asymmetries for charged pions (top), charged kaons (middle)
 and neutral kaons (bottom) on proton as a function of $x$, $z$ and $p^h_T$. }
\label{fig-call}    
\end{figure}
The results for charged kaons, although with larger statistical uncertainties,
show a similar trend: in particular the $K^+$ asymmetry has a negative trend
with increasing $x$, and the $K^-$ one is positive on average. The Collins
asymmetry for neutral kaons shows a positive trend with increasing $z$. The
average asymmetry is positive but compatible with zero within the statistical
uncertainty. In order to investigate in more detail the behaviour of the
asymmetries as a function of $z$ and $p_T^h$, the asymmetries for charged
hadrons were evaluated in a region where the signal is different from zero,
namely $x> 0.032$. The results are shown in Fig.~\ref{fig-cx32} for pions and
kaons. They are in good agreement with the other existing measurements on a
proton target from the HERMES experiment~\cite{ref:herm}. This is a non-obvious
result, as in the last $x$ bins the COMPASS $Q^2$ values are larger by a factor
3--4 than the HERMES ones. The weak $Q^2$ dependence of the Collins asymmetry is
also supported by a recent global fit~\cite{Anselmino:2013vqa} of the HERMES
pion results, the COMPASS preliminary pion asymmetries from the 2010 data, and
the Belle~\cite{belle} $e^+e^- \rightarrow \pi^+\pi^-$ asymmetries, which is
able to provide a good description of all the data sets. A comparison between
the final results of this paper and the fit is shown in Fig.~\ref{fig-pred}.

The Collins asymmetry for charged hadrons was further investigated by extending
the standard kinematic ranges in $z$ and $y$.  Compared to the above presented
results, the asymmetries extracted in the low-$z$ region ($0.1<z<0.2$) gave no
indication for a substantial $z$-dependence, neither for pions nor for kaons.
Similarly, in the low-y region ($0.05<y<0.1$) the pion asymmetries do not
exhibit any special behaviour, while the kaon ones suffer from too low
statistics.

\begin{figure}
\centering \includegraphics[width=0.6\textwidth,clip]{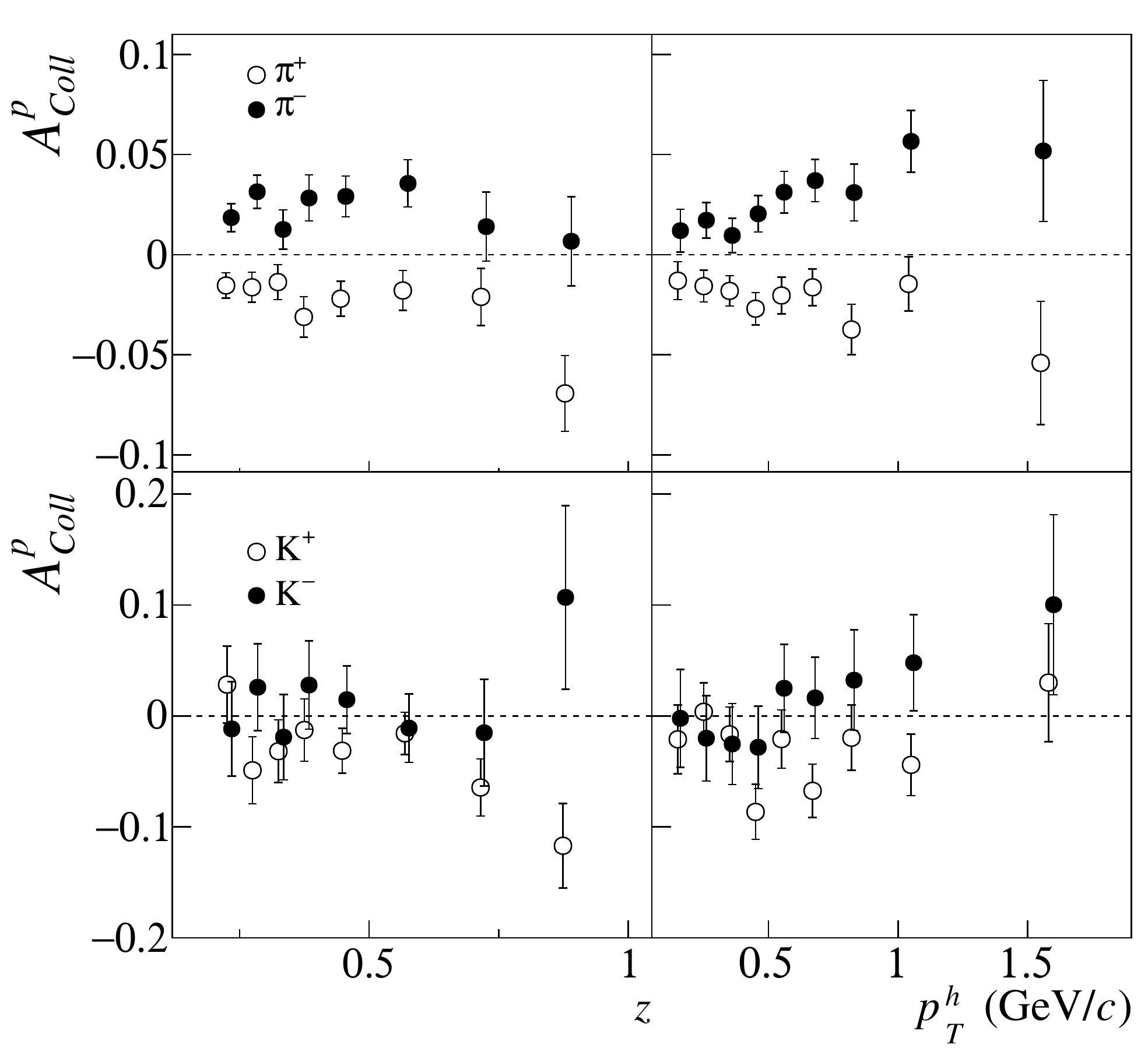}
\caption{The Collins asymmetries for pions (top) and kaons (bottom) as
 a function of $z$ and $p_T^h$, requiring $x>0.032$. }
\label{fig-cx32}    
\end{figure}
\begin{figure}
\centering \includegraphics[width=0.9\textwidth,clip]{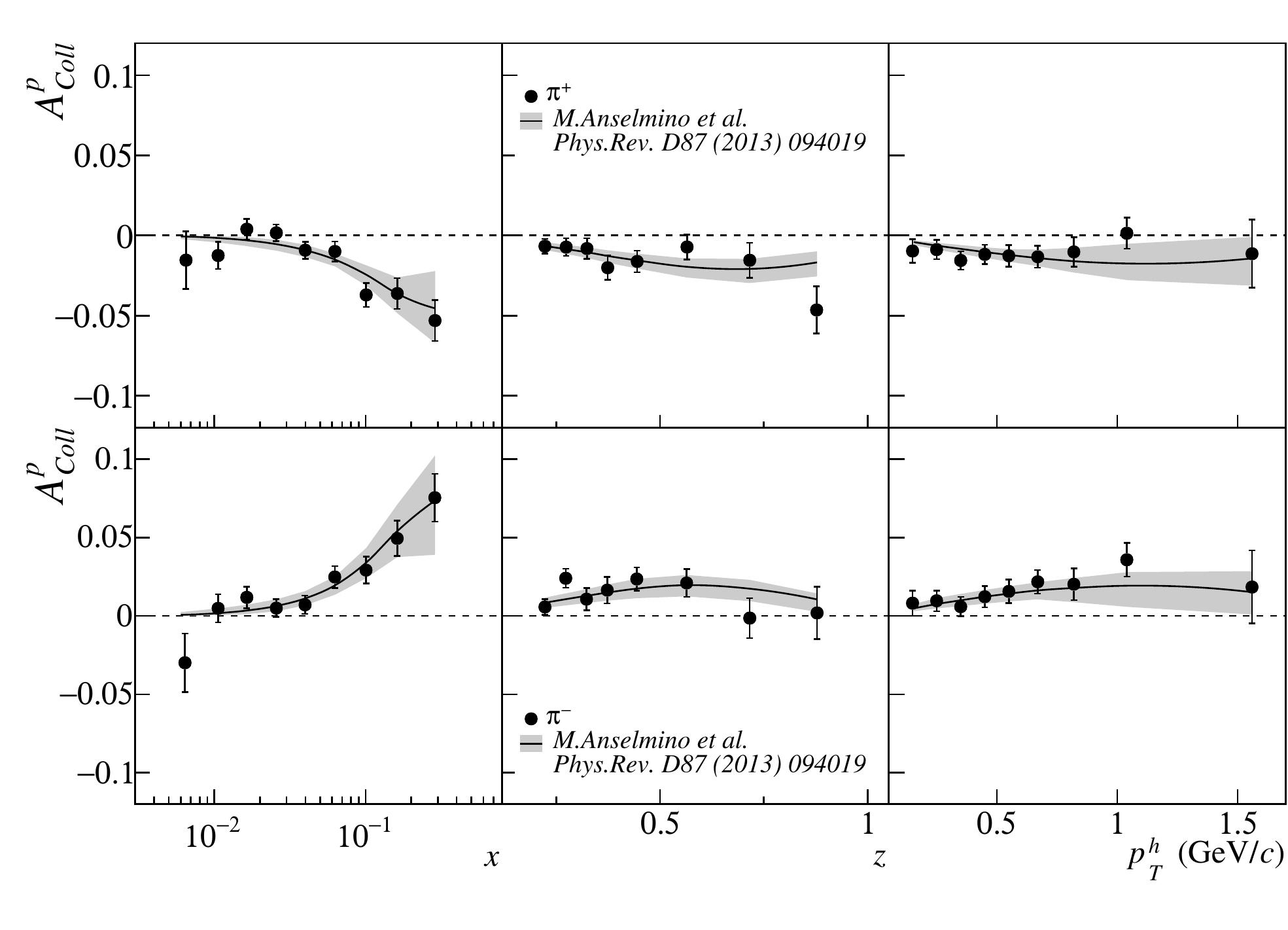}
\caption{Comparison between the Collins asymmetries for pions and one of the
 fits in~\cite{Anselmino:2013vqa} (fit with standard parameterisation and fit of
 $A_{12}$ Belle asymmetries~\cite{belle}). The preliminary asymmetries from
 2010 data are included in the fit. }
\label{fig-pred}    
\end{figure}

\begin{figure}
\centering \includegraphics[width=0.9\textwidth]{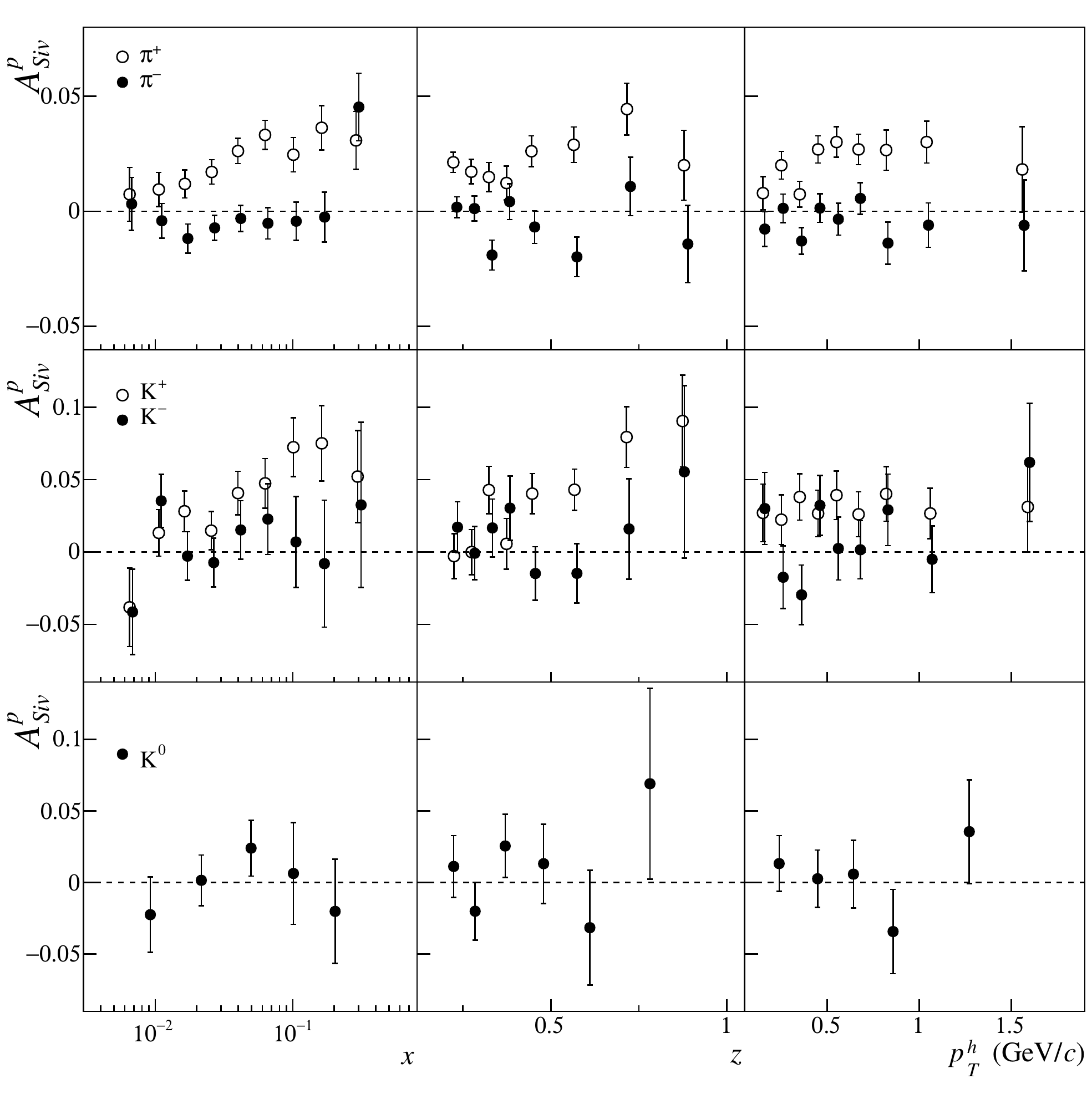}
\caption{The Sivers asymmetries for charged pions (top), charged kaons (middle)
 and neutral kaons (bottom) as a function of $x$, $z$ and $p_T^h$.}
\label{fig-2}    
\end{figure}
The Sivers asymmetries measured by COMPASS for pions and kaons on transversely
polarised protons are shown in Fig.~\ref{fig-2}. Also in this case, the pion
asymmetries are very similar to the unidentified hadron
asymmetries~\cite{ref:transvpsiv}. The asymmetries for negative pions and kaons,
as well as for neutral kaons are compatible with zero, while for positive pions
and kaons there is a clear evidence for a positive signal extending over the
full measured $x$ region and increasing with $z$. Very intriguing is the fact
that the $K^+$ signal is larger than the $\pi^+$ one, which indicates a possibly
not negligible role of sea
quarks~\cite{Efremov:2008vf,Efremov:2007kj,Anselmino:2008sga}. This is well
visible in Fig.~\ref{fig-comp}, where the two asymmetries are directly compared,
and from the mean values in the $x> 0.032$ region, which are respectively
$0.027\pm0.005$ and $0.043\pm0.014$. Unlike the case of the Collins asymmetry,
the Sivers asymmetry measured by COMPASS at large $x$ for positive pions and
kaons is smaller than the one from HERMES~\cite{ref:hermsiv}. This difference is
well visible also in the $z$ and $p_T^h$ variables when selecting the $x> 0.032$
region of the COMPASS data, as shown in Fig.~\ref{fig-sx32}.  Several fits,
which include the recently revisited $Q^2$ evolution, were performed using
HERMES asymmetries~\cite{ref:hermsiv}, COMPASS asymmetries on
deuteron~\cite{ref:transvdeut3} and for unidentified hadrons on
proton~\cite{ref:transvpsiv}. In Fig.~\ref{fig-preds}, the results of some of
these fits~\cite{fitsiv,fitsiv2,fitsiv3}, which employ $Q^2$ TMD evolutions, are
shown to well reproduce the COMPASS results. It will be interesting to see the
results of such fits when the results presented in this Letter will be included.
\begin{figure}
\centering \includegraphics[width=0.6\textwidth,clip=true]{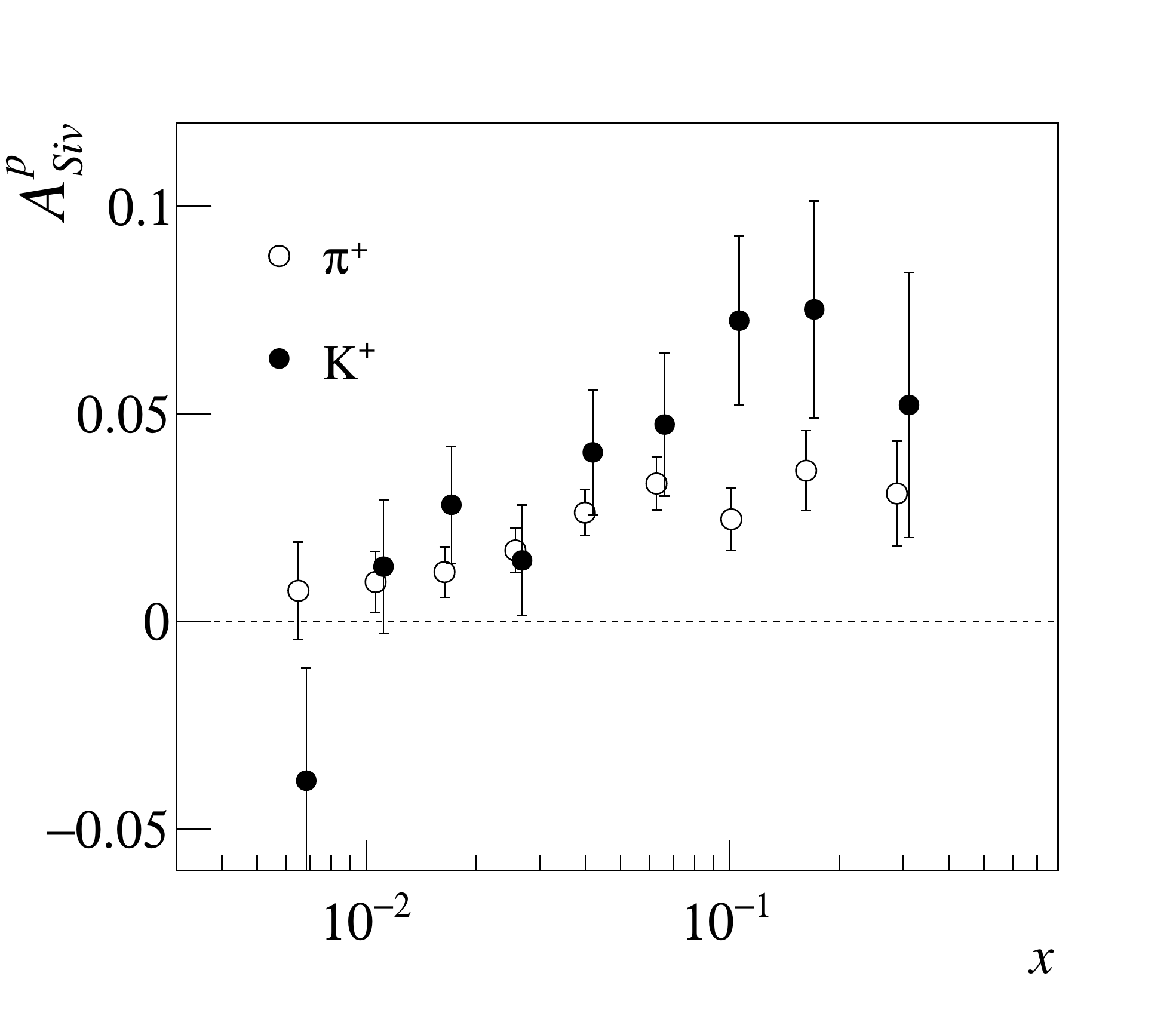}
\caption{The Sivers asymmetries for positive pions and kaons, as a function of
 $x$.}
\label{fig-comp}    
\end{figure}
\begin{figure}
\centering \includegraphics[width=0.9\textwidth,,trim=0 0 0 0,clip=true]{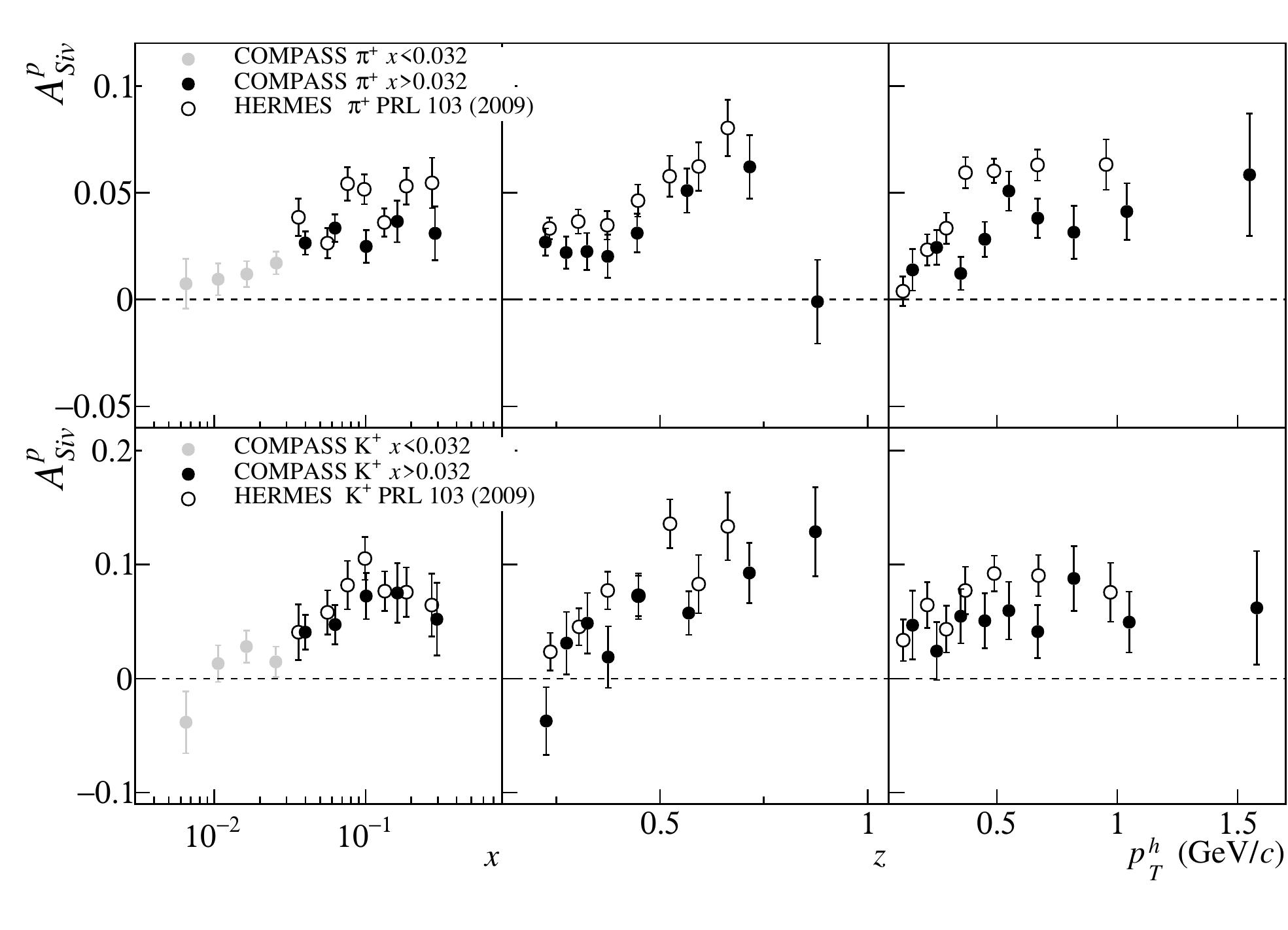}
\caption{The Sivers asymmetries for positive pions (top) and kaons (bottom) on
 proton as a function of $x$, $z$ and $p_T^h$, requiring $x> 0.032$. The
 asymmetries are compared to HERMES results. }
\label{fig-sx32}    
\end{figure}
\begin{figure}
\centering \includegraphics[width=0.9\textwidth,clip]{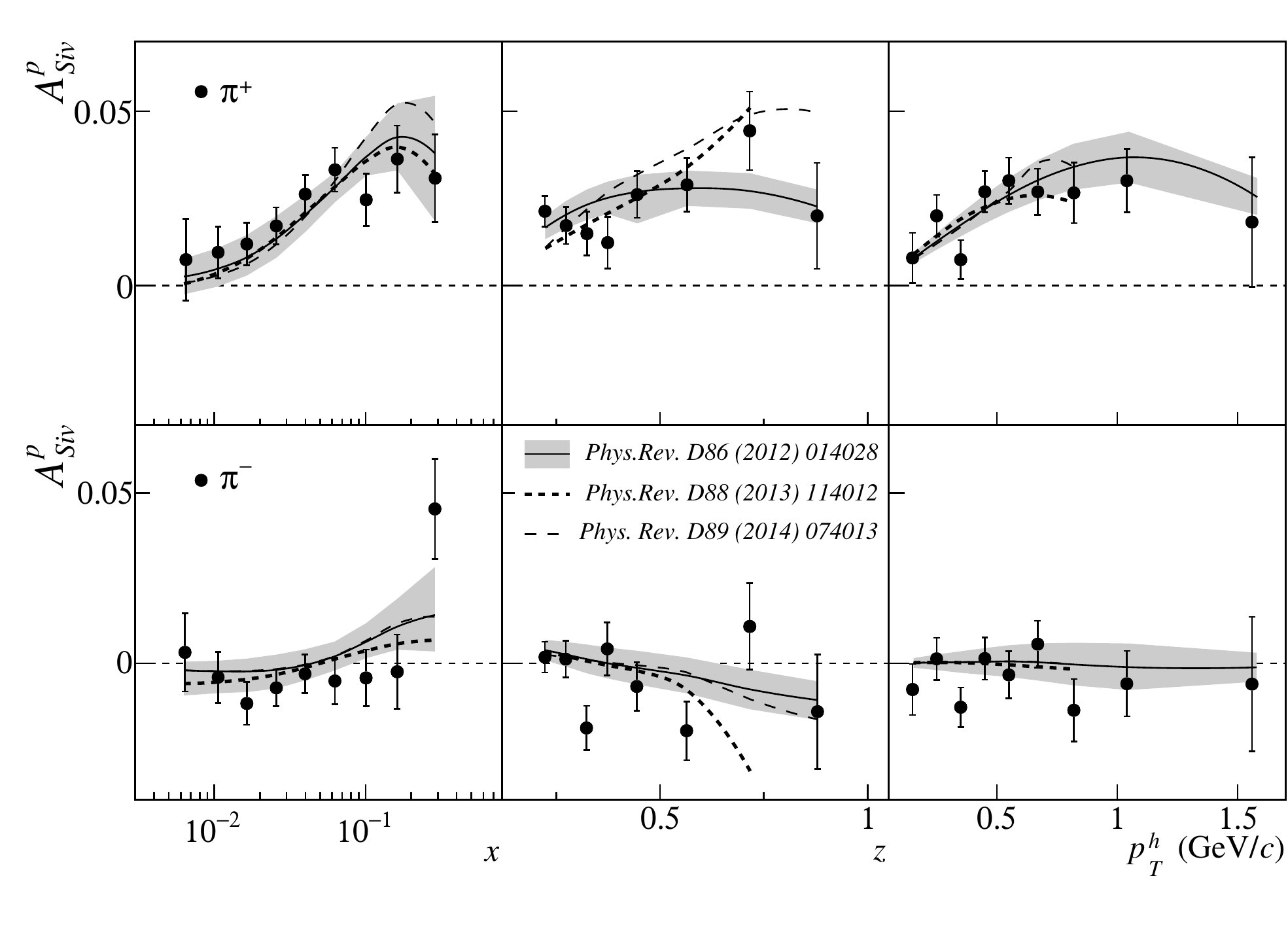}
\caption{Comparison between the Sivers asymmetries for pions and existing global
 fits~\cite{fitsiv,fitsiv2,fitsiv3}, in which the COMPASS results for the
 unidentified hadrons on protons~\cite{ref:transvpsiv} are included. }
\label{fig-preds}    
\end{figure}
More information on the $Q^2$ evolution is provided by the study of the Sivers
asymmetries in the low-$y$ region between 0.05 and 0.1. The pion asymmetries in
this region are compared in the left panel of Fig.~\ref{fig-slowz} to the
asymmetries obtained in the standard $y$ range and with the cut $x>0.032$. The
mean $Q^2$ values of these two samples are respectively $3.5\ (\mrf{GeV}/c)^2$ and 
$1.8\ (\mrf{GeV}/c)^2$. As for unidentified hadrons, there is an
indication for an increase of the $\pi^+$ asymmetries at low-$y$. The dependence of
the Sivers asymmetries with $z$ is further investigated considering the $z$
region between 0.1 and 0.2, where the asymmetries show smaller values. The
comparison of the pion asymmetries as a function of $x$ for the two separated $z$
ranges are shown in the right panel of Fig.~\ref{fig-slowz}. For negative pions,
a positive signal shows up in the low-$z$ region, which is not observed for
larger values of $z$. 
\begin{figure}
\centering 
\includegraphics[width=0.497\textwidth, clip=true]{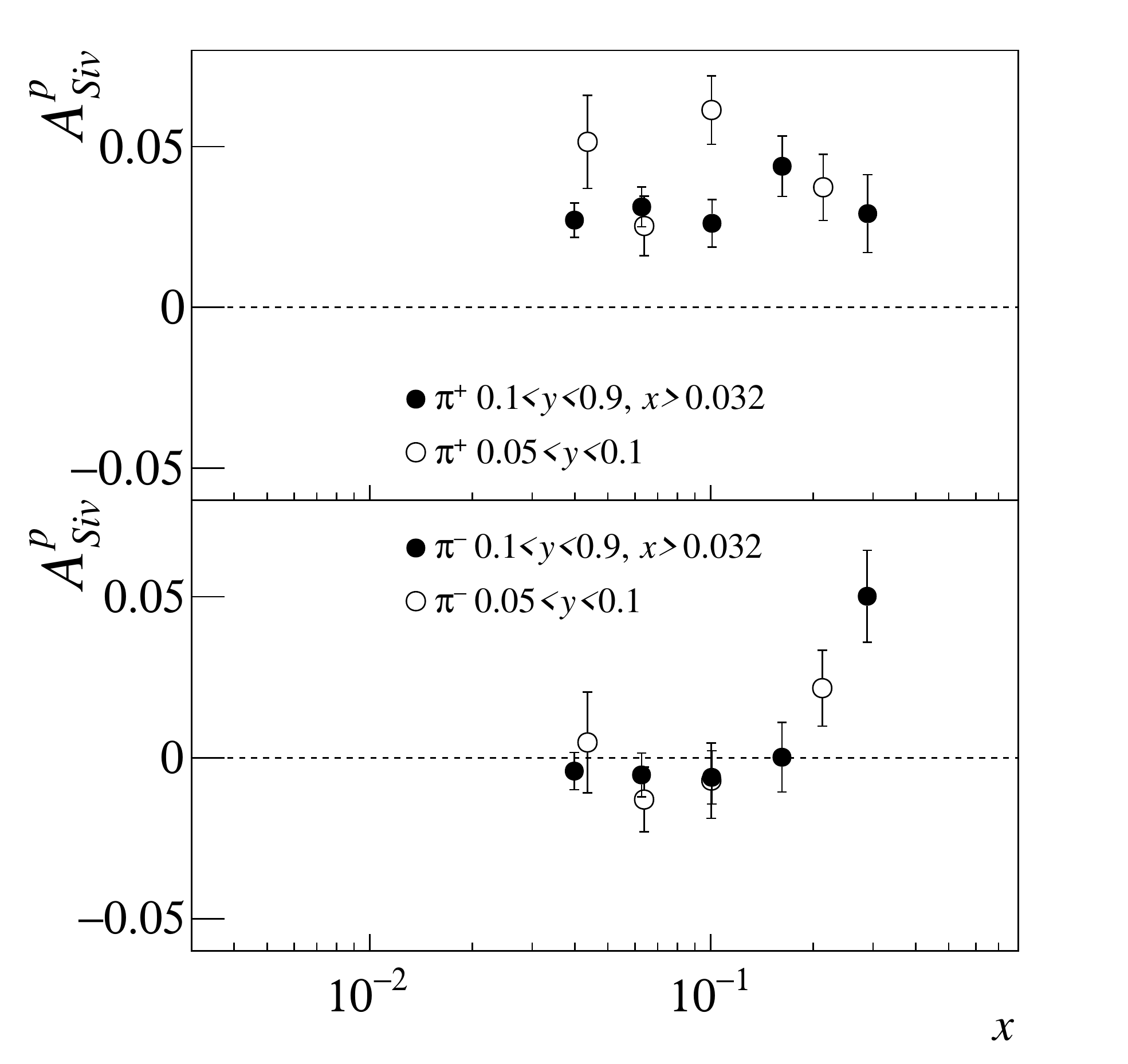}
\includegraphics[width=0.497\textwidth, clip=true]{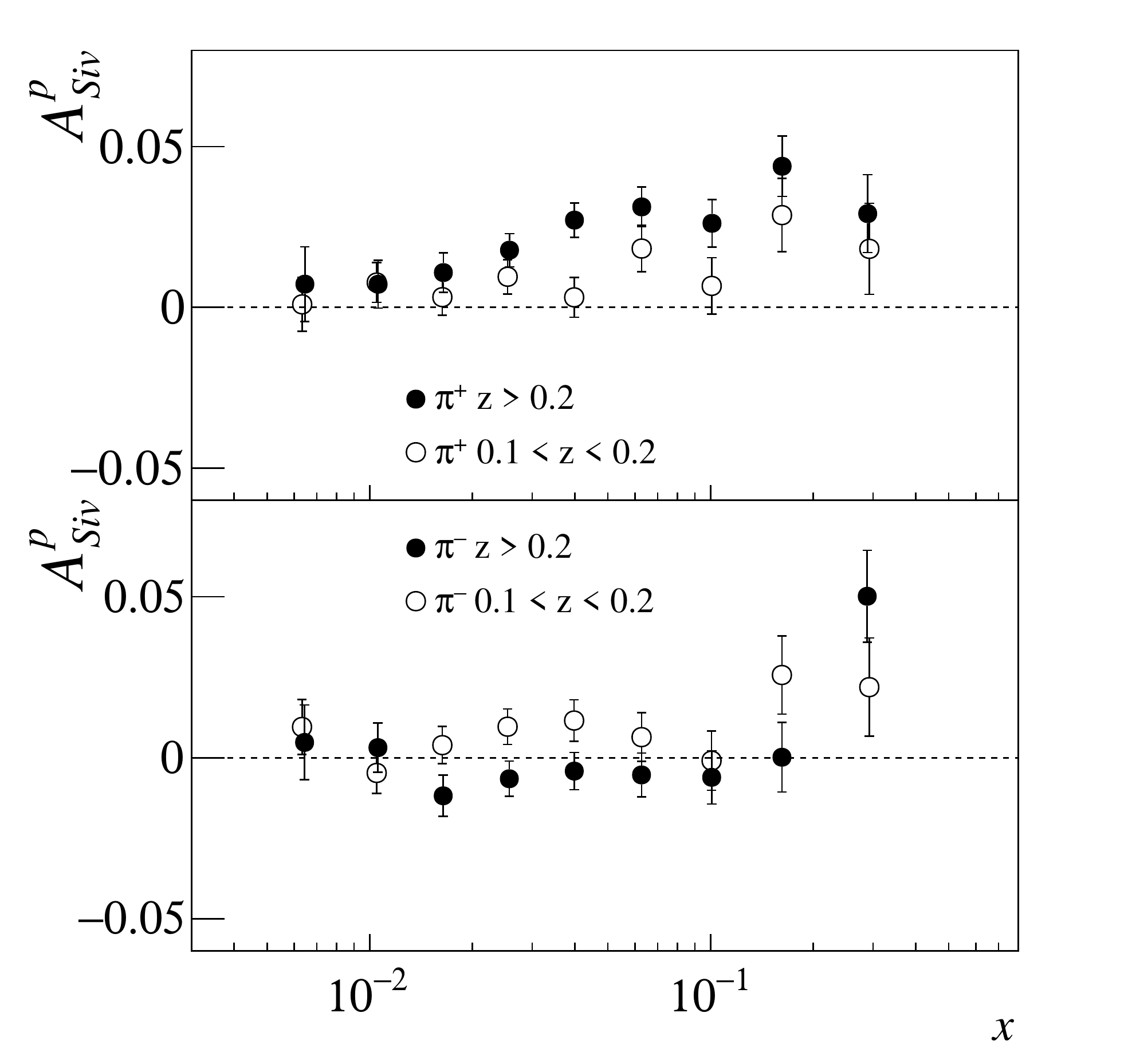}
\caption{The Sivers asymmetries for pions in different $y$ ranges (left) and $z$
 ranges (right), 2010 data.}
\label{fig-slowz}    
\end{figure}

In summary, using the high statistics data collected in 2007 and 2010, COMPASS
has measured the Collins and Sivers asymmetries in muonproduction of charged
pions and charged and neutral kaons produced off transversely polarised
protons. The high energy muon beam allowed the measurement of a broad kinematic
range in $x$ and $Q^2$. The $x$, $z$ and $p_T$ dependences of the asymmetries
were studied. Further investigations extending the range in $z$ and $y$ were
also performed. The Collins asymmetries are definitely different from zero for
pions and there are hints of a non-zero signal also for kaons, although in this
case the statistical significance is marginal. The Sivers asymmetries are
positive for positive pions and kaons, although different in size. This result
is of particular interest since it can be used to access the sea quark Sivers
PDFs. The results presented in this paper provide an important input for the
global analyses. Together with other measurements covering complementary
kinematic ranges, they allow the study of the $Q^2$ dependence of the asymmetries
and the quantitative extraction of the Collins FF and of the transversity and
Sivers PDFs. This information is crucial for the predictions for future
Drell-Yan measurements and for measurements at future high-energy electron-ion
colliders.

\noindent
\section* {Acknowledgements}
This work was made possible thanks to the financial support of our funding
agencies. We also acknowledge the support of the CERN management and staff, as
well as the skills and efforts of the technicians of the collaborating
institutes.

\end{document}